\documentclass[]{bytedance_seed}

\usepackage[toc,page,header]{appendix}
\usepackage{graphicx}
\usepackage{amsmath}
\usepackage{hyperref}
\usepackage{booktabs} 
\usepackage{caption}  
\usepackage{siunitx}  
\usepackage{multirow}
\usepackage{threeparttable} 
\usepackage{xcolor}

\usepackage{minitoc}

\title{Analytical Excited-State Gradients and Derivative Couplings in TDDFT with Minimal Auxiliary Basis Set Approximation and GPU Acceleration}
\author[1]{Zhichen Pu}
\author[1]{Xiaojie Wu}
\author[1]{Yuanheng Wang}
\author[1,2]{Cheng Fan}
\author[1]{Wen Yan}
\author[1,3,*]{Zehao Zhou}
\author[2,*]{Yi Qin Gao}
\author[1,*]{Qiming Sun}

\affiliation[1]{ByteDance Seed}
\affiliation[2]{New Cornerstone Science Laboratory, College of Chemistry and Molecular Engineering, Peking University, Beijing 100871, China}
\affiliation[3]{Zhongguancun Academy}
\contribution[*]{Corresponding authors}

\abstract{
Calculating excited-state gradients and derivative couplings using time-dependent density functional theory (TDDFT) remains a computationally demanding task.
An efficient variant, TDDFT with resolution of the identity and a minimal auxiliary basis (TDDFT-ris), has been developed to accelerate excitation energy calculations.
However, the formulation and implementation of analytical derivatives for this method have not yet been reported.
In this work, we present an implementation of analytical excited-state gradients and derivative couplings within the TDDFT-ris framework.
Benchmark calculations on medium-sized organic molecules demonstrate a two- to three-fold speedup for both gradients and derivative couplings compared to standard TDDFT.
The accuracy of the TDDFT-ris approach is assessed for gradient-dependent applications, including geometry optimizations, emission energy calculations, and the localization of minimum-energy crossing points.
Overall, the TDDFT-ris method provides reliable approximations for most cases, with noticeable errors mainly occurring in derivative couplings between nearly degenerate states.
}

\date{\today}
\correspondence{Qiming Sun at \email{qiming.sun@bytedance.com}, Zehao Zhou at
\email{zhouzehao@bjzgca.edu.cn}, Yiqin Gao at \email{gaoyq@pku.edu.cn}}

\begin{document}
\maketitle


\section{Introduction}

Excited states play a crucial role in the study of photochemistry, electronic spectroscopy, non-adiabatic processes, electronic transitions, energy transfer, and charge transfer problems. 
The computation of excited-state gradients within time-dependent density functional theory (TDDFT) is crucial for excited-state geometry optimization, enabling prediction of fluorescence emission wavelengths in molecules. 
Similarly, derivative couplings (also known as first-order non-adiabatic coupling matrix elements, fo-NACMEs) are essential for simulating molecular dynamics (MD) on excited state potential energy surfaces, particularly when coupled with Fewest Switches Surface Hopping (FSSH) to explore critical regions, such as conical intersections that govern non-radiative decay pathways\cite{tully1990molecular,herbert2016beyond,tully2012perspective}.


The analytical derivatives of excited states can be computed using various wavefunction-based methods, including the equation-of-motion coupled-cluster theory\cite{gauss2002analytic,stanton1994analytic,tajti2009analytic,faraji2018calculations}, the algebraic diagrammatic construction (ADC) method\cite{rehn2019analytic,mai2017surface}, selected configuration interaction (CI)\cite{coe2023analytic,coe2023analytic2}, state-average multi-configurational self-consistent field methods\cite{lengsfield1984evaluation,freitag2019approximate} and multi-reference CI methods\cite{lischka2004analytic,fdez2016analytical}.
However, these methods are considered computationally demanding and are often unaffordable for medium-sized molecules.
Density Functional Theory (DFT)\cite{kohn1965self, hohenberg1964inhomogeneous} has revolutionized computational chemistry and materials science by providing an efficient and accurate framework to describe the electronic structure of molecules and solids.
The Time-Dependent Density Functional Theory (TDDFT) \cite{runge1984density} extends DFT to describe electronic excited states.
Compared to the wavefunction theories, TDDFT offers a reasonable estimation of excited state energies and analytical derivatives at a more affordable computational cost.
The calculation of TDDFT gradients and derivative couplings has been extensively investigated\cite{furche2002adiabatic,rappoport2005analytical,scalmani2006geometries,tapavicza2007trajectory,ou2015first,herbert2016beyond,zhang2014analytic,zhang2015analytic,send2010first,li2014first,subotnik2015requisite,zhang2021nonadiabatic,uratani2020fast}.
Analytical derivative couplings within the TDDFT framework were developed separately by Li and Zhang, using time-dependent response theory to compute the excited state derivative couplings\cite{li2014first,li2014first2,zhang2014analytic,zhang2015analytic,send2010first,herbert2022spin}.
Qi et al. also derived similar TDDFT equations based on the time-dependent many-body wavefunction framework \cite{ou2015first}.

Although TDDFT significantly improves computational efficiency compared to wavefunction-based methods, TDDFT-based MD are still limited to relatively small molecules \cite{herbert2016beyond,subotnik2016understanding,muller2025machine} due to the substantial computational cost of evaluating gradients and derivative couplings.
Applying TDDFT derivative computations to large-scale systems with hundreds or thousands of atoms, such as conjugated polymers, biomolecular chromophores, or nanostructured materials, is not feasible on a routine basis.
Furthermore, the MD simulation with FSSH schemes requires hundreds of trajectories and thousands steps per trajectory to ensure statistically robust results\cite{miller2024ultrafast,parker2020surface}, further increasing the computational demands.
Consequently, various approximations and adjustments to the nuclear motion have been employed to estimate the coupling between electrons and nuclei\cite{Sangiogo-Gil2025,Ryabinkin2015,diaz2024non,bannwarth2019gfn2,Wu2022}.

The primary computational bottleneck in the TDDFT method lies in the calculation of the electron-repulsion integrals within the linear-response kernel.
To address this computational challenge, various approximation methods have been developed to balance accuracy and computational efficiency \cite{walker2006efficient,ruger2016tight, wu2011linear,rocca2008turbo,zuehlsdorff2013linear,giannone2020minimal,baseggio2015new,niehaus2001tight,niehaus2009approximate,grimme2013simplified,bannwarth2014simplified,DAntoni2025}.
Some of these approaches have been extended to excited-state gradients \cite{havenridge2023analytical} and derivative couplings \cite{niehaus2021ground,wu2022nonadiabatic,niehaus2023exact}.
Semi-empirical two-electron integrals have been used to approximate the linear-response kernel, leading to the development of simplified TDDFT (sTDDFT/sTDA) methods\cite{grimme2013simplified, de2024exact}.
The minimal auxiliary basis model (TDDFT-as) was proposed to approximate the Coulomb matrix in the linear-response kernel for semi-local density functionals \cite{giannone2020minimal}.
In this approach, a single $s$-type Gaussian auxiliary function is used per atom.
This minimal auxiliary basis model was extended to the computation of the exchange matrix for the hybrid exchange-correlation functionals by Zhou, introducing the TDDFT-ris method \cite{zhou2023minimal,zhou2024converging}.
The TDDFT-ris method achieves a speedup of up to two orders of magnitude compared to conventional TDDFT without compromising accuracy.

Another effective strategy for improving TDDFT computational performance is GPU acceleration \cite{Kim2023,Kim2024}.
GPU architectures are particularly well suited for tensor operations, making them highly compatible with density-fitting approaches\cite{weigend2008hartree}.
The density-fitting implementation in GPU4PySCF achieves performance improvements of two to three orders of magnitude over conventional CPU implementations for DFT energies, analytical nuclear gradients, and Hessians\cite{wu2025enhancing,li2025introducing}.
The analytical-derivative module in GPU4PySCF provides general APIs that can be readily extended to excited-state derivative calculations.

Building upon this foundation, this work extends the TDDFT-ris method to compute analytical nuclear gradients and derivative couplings.
We will present the theoretical background of TDDFT, the TDDFT-ris method, and their analytical derivatives in Section~\ref{sec:theory}.
We will then briefly discuss how the features of standard TDDFT and the TDDFT-ris approximation are implemented with GPU support in Section~\ref{sec:implementation}.
In Section~\ref{sec:benchmark}, we evaluate the performance of the TDDFT-ris approximation to determine the performance improvements it offers over standard TDDFT computations.
In Section~\ref{sec:application}, we assess the accuracy of TDDFT-ris gradients and derivative couplings using a series of metrics, including the optimization of excited-state geometries, the transition probabilities between electronic states, and the minimum-energy crossing point (MECP).

\section{Theory}
\label{sec:theory}

In this section, we briefly review the theoretical framework of TDDFT and its extension to the analytical calculation of excited-state gradients and derivative couplings.
From this foundation, we derive the TDDFT-ris approximation for the derivative calculations.

\subsection{TDDFT gradients and analytical derivative couplings}

Within the linear-response TDDFT framework for closed-shell systems, as formulated by Casida\cite{casida1995time}, the excitation energies are obtained by solving the following generalized eigenvalue problem:
\begin{align}
    \begin{pmatrix}
        \boldsymbol{A} & \boldsymbol{B} \\
        \boldsymbol{B} & \boldsymbol{A}
        \end{pmatrix}
        \begin{pmatrix}
        \boldsymbol{X}_I \\
        \boldsymbol{Y}_I
        \end{pmatrix}
        =&
        \omega_I
        \begin{pmatrix}
        \boldsymbol{1} & \boldsymbol{0} \\
        \boldsymbol{0} & -\boldsymbol{1}
        \end{pmatrix}
        \begin{pmatrix}
        \boldsymbol{X}_I \\
        \boldsymbol{Y}_I
    \end{pmatrix}, \label{equ:casida} \\
    A_{ai, bj} =& (\varepsilon_a - \varepsilon_i) \delta_{ab} \delta_{ij} + K_{ai, bj}, \label{equ:A_matrix} \\
    B_{ai, bj} =& K_{ai, jb}, \label{equ:B_matrix} \\
    K_{pq,rs} =& g_{pq,sr} + (f_\text{xc})_{pq,sr}, \label{equ:K_matrix} \\
    g_{pq,sr} =& (pq|sr) - c_x(pr|sq) \label{equ:g_integral}, \\
    (f_\text{xc})_{pq,sr} = & \frac{\partial^2 E_{\text{xc}}}{\partial D_{qp} \partial D_{rs}}, \label{equ:g_xc} \\
    (pq|sr) = & \iint \mathrm{d} \boldsymbol{r}_1 \mathrm{d} \boldsymbol{r}_2  \frac{p(\boldsymbol{r}_1) q(\boldsymbol{r}_1) s(\boldsymbol{r}_2) r(\boldsymbol{r}_2)}{|\boldsymbol{r}_1 - \boldsymbol{r}_2|}  , \label{equ:pqrs}
\end{align}
where the eigenvalue for state $I$ is $\omega_I$, and the eigenvector is $\begin{pmatrix}\boldsymbol{X}_I \\\boldsymbol{Y}_I\end{pmatrix}$. 
In Eq.~\eqref{equ:A_matrix}, $ \varepsilon $ is the orbital energy, and $ K_{pq,rs} $ is the element of the coupling matrix.
In this manuscript, we use $a,b$ to denote virtual orbitals, $i,j$ denote occupied orbitals, and $p,q,r,s$ denote dummy orbitals, which can refer to either molecular orbitals or atomic orbitals.
The four-center two-electron repulsion integral is denoted as $(pq|sr)$ as defined in Eq.~\eqref{equ:pqrs}.
The integral $ - c_x(pr|sq) $ from Eq.~\eqref{equ:g_integral} together with $ (f_\text{xc})_{pq,rs} $ in Eq.~\eqref{equ:K_matrix} jointly form the exchange-correlation (XC) functional kernel, where $ c_x $ presents the fraction of exact exchange contribution. 
The pure functional component, $ (f_\text{xc})_{pq,rs} $, is the second-order derivative of the XC functional $ E_{\text{xc}} $ with respect to the density matrix $ \boldsymbol{D} $, as defined in Eq.~\eqref{equ:g_xc}.


The theoretical framework for calculating gradients and derivative couplings within TDDFT is well-established\cite{furche2002adiabatic,rappoport2005analytical,scalmani2006geometries,tapavicza2007trajectory,ou2015first,herbert2016beyond,zhang2015analytic,send2010first,li2014first,subotnik2015requisite}.
Following the unified formalism developed by Li and co-workers\cite{li2014first,li2014first2,wang2021nac},
the derivative coupling between two electronic states, $\Psi_I$ and $\Psi_J$, is defined as:
\begin{equation}
    \boldsymbol{g}_{IJ}^{\xi} = \langle \Psi_I | \frac{\partial}{\partial \xi} | \Psi_J \rangle, \quad \xi \in x = \{\boldsymbol{R}_1, \boldsymbol{R}_2, \ldots, \boldsymbol{R}_N\},
    \label{eq:fo-nacme}
\end{equation}
where $\xi$ is a nuclear coordinate of the $N$-atom molecule.
The indices $I$ and $J$ represent specific electronic states. 
When $I$ is set to 0, it denotes the ground state. The gradient for state $I$ is defined as
\begin{equation}
  \boldsymbol{g}_{I}^{\xi} = \frac{\partial}{\partial \xi} E_I,
  \label{eq:gradient}
\end{equation}
where $E_I$ is the energy of the $I$-th electronic state.

The derivative computation must account for the response of molecular orbitals with respect to the nuclear coordinates.
To incorporate this constraint, a Lagrangian\cite{furche2001density, furche2002adiabatic} is introduced as follows:
\begin{align}
    L[x, \boldsymbol{C}(x), \boldsymbol{Z}, \boldsymbol{W}] = & g[x, \boldsymbol{C}(x)] + \sum_{ai} Z_{ai} F_{ai}(x)
    - \sum_{pq} W_{pq} (S_{pq}(x) - \delta_{pq}), 
    \label{equ:Lagrangian}
\end{align}
where $\boldsymbol{C}(x)$ represents the molecular orbital coefficients at a given atomic geometry $x$.
Two constraints are enforced via the Lagrange multipliers, $\boldsymbol{Z}$ and $\boldsymbol{W}$.
The constraint $F_{ai}(x) = 0$, associated with $\boldsymbol{Z}$, corresponds to the stationary conditions of the ground state energy w.r.t. molecular orbital coefficients:
\begin{equation}
  F_{ai}(x) = \frac{\partial E_\text{GS}}{\partial C_{ai}} = 0.
\end{equation}
The other constraint, $S_{pq}(x) - \delta_{pq} = 0$, ensures the orthonormality of the molecular orbitals.

The core of the Lagrangian~\eqref{equ:Lagrangian} is the generating function $g[x, \boldsymbol{C}(x)]$.
Its specific form depends on the property to be calculated\cite{li2014first2,li2014first}, and these terms are well-discussed in Ref.~\citenum{li2014first}. 
For convenience, we list the generating functions here. 
For the excited-state gradient in TDDFT, it takes the following form:
\begin{align}
    g_I[x, \boldsymbol{C}(x)] = & \langle \boldsymbol{F}; \boldsymbol{T}_{II} \rangle + \langle \boldsymbol{g}; \boldsymbol{\Gamma}_{II} \rangle + \langle \boldsymbol{f}_{\text{xc}}; \{ \boldsymbol{R}_I^S, \boldsymbol{R}_I^S \} \rangle. \label{equ:gI}
\end{align}
For the derivative couplings between the ground state and an excited state, or between two excited states, the corresponding forms are
\begin{align}
    g_{0I}[x, \boldsymbol{C}(x)] = & \langle \underline{\boldsymbol{d}}; \underline{\boldsymbol{\gamma}}^{0I} \rangle, \label{equ:g0I} \\
    g_{IJ}[x, \boldsymbol{C}(x)] = & (E_J - E_I)^{-1}\left(\langle\boldsymbol{F}; \boldsymbol{T}_{IJ} \rangle + \langle \boldsymbol{g}; \boldsymbol{\Gamma}_{IJ} \rangle + \langle \boldsymbol{f}_{\text{xc}}; \{ \boldsymbol{R}_I^S, \boldsymbol{R}_J^S \} \rangle\right) 
    + \langle \underline{\boldsymbol{d}}; \underline{\boldsymbol{\gamma}}^{IJ} \rangle. \label{equ:gIJ}
\end{align}
We follow the notations of Ref.~\citenum{li2014first} for the matrices and symbols used in Eqs.~\eqref{equ:gI} to \eqref{equ:gIJ}, which are also summarized in the Appendix~\ref{subsec:appendix_generating} for quick reference. It should be noted that the quadratic-response term is neglected, following the recommendation in Ref.~\citenum{wang2021nac}.

Finally, we directly list the final expressions for the excited-state gradient and derivative couplings between the ground state and excited states, or between excited states
\begin{align}
  g_I^{\xi} &= \langle \boldsymbol{H}^{(\xi)}; \boldsymbol{P}_I \rangle + \langle \boldsymbol{g}^{(\xi)}; \{ \boldsymbol{D}, \boldsymbol{P}_I \} + \boldsymbol{\Gamma}_{II} \rangle \nonumber\\
  & \quad + \langle \boldsymbol{v}_{\text{xc}}^{(\xi)}; \boldsymbol{P_I} \rangle + \langle \boldsymbol{f}_{\text{xc}}^{(\xi)}; \{ \boldsymbol{R}_I^S, \boldsymbol{R}_I^S \} \rangle 
  - \langle \boldsymbol{S}^{(\xi)}; \boldsymbol{W} \rangle. \label{equ:gI_xi} \\
  g_{0I}^{\xi} &= \langle \boldsymbol{H}^{(\xi)}; \boldsymbol{Z}^S \rangle + \langle \boldsymbol{g}^{(\xi)}; \{ \boldsymbol{D}, \boldsymbol{Z}^S \} \rangle 
  + \langle \boldsymbol{v}_{\text{xc}}^{(\xi)}; \boldsymbol{Z}^S \rangle - \langle \boldsymbol{S}^{(\xi)}; \boldsymbol{W} \rangle + \langle \boldsymbol{d}^{(\xi)}; \boldsymbol{\gamma}^{0I} \rangle.\\
  g_{IJ}^{\xi} & = (E_J-E_I)^{-1} \tilde{L}_{IJ}^{(\xi)}, \\
  \tilde{L}_{IJ}^{(\xi)}  &= \langle \boldsymbol{H}^{(\xi)}; \boldsymbol{P}_{IJ} \rangle + \langle \boldsymbol{g}^{(\xi)}; \{ \boldsymbol{D}, \boldsymbol{P}_{IJ} \} + \boldsymbol{\Gamma}_{IJ} \rangle \nonumber \\
  &\quad + \langle \boldsymbol{v}_{\text{xc}}^{(\xi)}; \boldsymbol{P}_{IJ} \rangle + \langle \boldsymbol{f}_{\text{xc}}^{(\xi)}; \{ \boldsymbol{R}_I^S, \boldsymbol{R}_J^S \} \rangle
  - \langle \boldsymbol{S}^{(\xi)}; \boldsymbol{\tilde{W}} \rangle + \langle \boldsymbol{d}^{(\xi)}; \boldsymbol{\tilde{\gamma}}^{IJ} \rangle. \label{equ:gIJ_xi}
\end{align}
The matrix $ \boldsymbol{Z} $ is determined by solving the so-called Z-vector equation\cite{handy1984evaluation}
\begin{align}
    (\boldsymbol{A} + \boldsymbol{B}) \boldsymbol{Z} = g^{(\mathrm{VO})} - g^{(\mathrm{OV})}. \label{equ:Z_vector}
\end{align}
The notations used in Eqs. \eqref{equ:gI_xi} to \eqref{equ:Z_vector} are defined in Appendix \ref{subsec:appendix_derivatives}. 
The right-hand side of the $Z$-vector equation~\eqref{equ:Z_vector} depends on the specific property being calculated.
For further details, the reader is referred to Ref.~\citenum{li2014first}.

\subsection{Excited-state gradient and derivative couplings for TDDFT-ris}

The most computationally demanding step in solving the TDDFT Casida equation~\eqref{equ:casida} is the evaluation of the coupling matrix $\boldsymbol{K}$ \eqref{equ:K_matrix}.
It involves the evaluation of two-electron integrals and the numerical integration of the XC kernel.
To reduce the computational cost of the coupling matrix evaluation, the TDDFT-ris method employs the resolution of identity (RI) with a small auxiliary basis set to approximate the two-electron integrals within the coupling matrix.
Additionally, this method neglects the contribution from the pure XC functional kernel, \textit{i.e.},
\begin{align}
    K_{pq,rs} \approx & K_{pq,rs}^\text{\text{ris}}  =  g^\text{\text{ris}}_{pq,sr}, \label{equ:ris} \\
    g^\text{\text{ris}}_{pq,sr} = & \sum_{AB}(pq|A)(\boldsymbol{M}^{-1})_{AB}(B|sr)
    - c_x \sum_{AB}(pr|A)(\boldsymbol{M}^{-1})_{AB}(B|sq), \label{equ:g_modified} \\
    M_{AB} &= (A|B),
\end{align}
where $A$ and $B$ denote auxiliary basis functions, while $(A|B)$ and $(pq|A)$
represent the two-center and three-center two-electron repulsion integrals, respectively.

The auxiliary basis set comprises a single Gaussian-type orbital (GTO) for each angular momentum component (up to $l=2$) on each atom.
The use of a minimal auxiliary basis set for the RI approximation introduces larger errors compared to the regular density fitting approximation with optimized JK-fit auxiliary basis sets.
In the TDDFT-ris scheme, distinct auxiliary basis sets can be employed for the Coulomb (J-fit) and exchange (K-fit) integrals, corresponding to the first and second terms in Eq.~\eqref{equ:g_modified}, respectively. 
The exponents of these auxiliary GTOs are determined by minimizing the deviation of the resulting total excitation energies from standard TDDFT reference values~\cite{zhou2023minimal}.

Within the TDDFT-ris formalism, only the generating functions in Eq. \eqref{equ:Lagrangian} are modified as follows:
\begin{align}
  g_I = & \langle \boldsymbol{F}; \boldsymbol{T}_{II} \rangle + \langle \boldsymbol{g}^{\text{ris}}; \boldsymbol{\Gamma}_{II} \rangle, \\
  g_{IJ} = & (E_J - E_I)^{-1} \left(\langle \boldsymbol{F}; \boldsymbol{T}_{IJ} \rangle + \langle \boldsymbol{g}^{\text{ris}}; \boldsymbol{\Gamma}_{IJ} \rangle\right) + \langle \underline{\boldsymbol{d}}; \underline{\boldsymbol{\gamma}}^{IJ} \rangle. 
\end{align}
In contrast, the generating functions for the derivative couplings between the ground and excited states remain identical to those in standard TDDFT~\eqref{equ:g0I}. The excited-state gradient and derivative couplings are computed as
\begin{align}
  g_I^\xi
  = & \langle \boldsymbol{H}^{(\xi)}; \boldsymbol{P}_I \rangle 
  + \langle \boldsymbol{g}^{(\xi)}; \{ \boldsymbol{D}, \boldsymbol{P}_I \} \rangle 
  + \langle \boldsymbol{g}^{\text{ris},(\xi)}; \boldsymbol{\Gamma}_{II} \rangle
  + \langle \boldsymbol{v}_{\text{xc}}^{(\xi)}; \boldsymbol{P}_I \rangle 
  - \langle \boldsymbol{S}^{(\xi)}; \boldsymbol{W} \rangle \label{equ:gI_xi_ris}, \\
  g_{IJ}^{\xi} 
  = & (E_J - E_I)^{-1}  \left( \langle \boldsymbol{H}^{(\xi)}; \boldsymbol{P}_{IJ} \rangle 
  + \langle \boldsymbol{g}^{(\xi)}; \{ \boldsymbol{D}, \boldsymbol{P}_{IJ} \} \rangle
  + \langle \boldsymbol{g}^{\text{ris},(\xi)}; \boldsymbol{\Gamma}_{IJ} \rangle
  + \langle \boldsymbol{v}_{\text{xc}}^{(\xi)}; \boldsymbol{P}_{IJ} \rangle\right. \nonumber \\
  & \left. - \langle \boldsymbol{S}^{(\xi)}; \boldsymbol{\tilde{W}} \rangle + \langle \boldsymbol{d}^{(\xi)}; \boldsymbol{\tilde{\gamma}}^{IJ} \rangle \right). \label{equ:gIJ_xi_ris}
\end{align}
It should be noted that the RI approximation in TDDFT-ris does not modify every term that involves two-electron integrals or exchange–correlation (XC) kernel integration.
The right-hand side of the $Z$-vector equation~\eqref{equ:Z_vector} within the TDDFT-ris framework is updated according to the specific property being calculated (see Appendix~\ref{subsec:appendix_derivatives_ris} for details).
The left-hand side of the $Z$-vector equation is identical to that in standard TDDFT, as these tensors originate from the orbital response of the ground-state calculation.

To mitigate errors in the TDDFT-ris scheme, distinct auxiliary basis sets can be utilized for the J-fit and K-fit terms.
Upon extending the TDDFT-ris method to the calculation of analytical gradients and derivative couplings, certain derivative integrals $\boldsymbol{g}^{(\xi)}$ in Eqs.~\eqref{equ:gI_xi} and \eqref{equ:gIJ_xi} are replaced by their TDDFT-ris counterparts, $\boldsymbol{g}^{\text{ris},(\xi)}$, as shown in Eqs.~\eqref{equ:gI_xi_ris} and \eqref{equ:gIJ_xi_ris}.
The derivatives of the two-electron integrals are evaluated as follows:
\begin{align}
  \frac{\partial}{\partial \xi} \left[ \sum_{AB}(pq|A)(\boldsymbol{M}^{-1})_{AB}(B|sr)\right] = & \sum_{AB}(pq|A)^{\xi}(\boldsymbol{M}^{-1})_{AB}(B|sr)  + \sum_{AB}(pq|A)(\boldsymbol{M}^{-1})_{AB}(B|sr)^{\xi} \nonumber \\
  &- \sum_{ABCD}(pq|A)(\boldsymbol{M}^{-1})_{AC}\boldsymbol{M}^{\xi}_{CD}(\boldsymbol{M}^{-1})_{DB}(B|sr).
\end{align}

\section{Implementation}\label{sec:implementation}

The program for calculating TDDFT gradients and derivative couplings for both standard TDDFT and the approximate TDDFT-ris methods has been developed within the GPU4PySCF package\cite{sun2020recent,li2025introducing,wu2025enhancing,lehtola2018recent,pu2025enhancing}.
In this TDDFT derivative program, we have integrated various methods for calculating two-electron integrals and their gradients, including density fitting approximations and exact analytical integral evaluations.
Two-electron integrals and their derivatives are evaluated using the GPU4PySCF integral implementations\cite{li2025introducing,wu2025enhancing}.
The high-order derivatives of the XC functionals are evaluated using the LibXC library\cite{Lehtola2018}.
We will not discuss the details of these integral evaluation algorithms in this work, as they are beyond the scope of this study.
The source code for excited-state gradients and derivative couplings for both TDDFT and TDDFT-ris is publicly available in the GPU4PySCF GitHub repository at \url{https://github.com/pyscf/gpu4pyscf}. 
Our implementation is validated via finite-differences as shown in Appendix \ref{app:benchmark_test}.

To simplify the code management across multiple TDDFT variants, our current program implementation integrates TDDFT and its counterpart within the Tamm-Dancoff Approximation \cite{hirata1999time} (TDA) into a unified code framework.
In this setup, TDA gradients and derivative couplings are evaluated using the TDDFT routines by setting the $\boldsymbol{Y}$ amplitudes to zero.
A similar hierarchical structure and implementation strategy is also employed for the TDDFT-ris and TDA-ris methods.

The TDDFT and its derivative code is initially implemented using exact integral computation routines.
Subsequently, density fitting with regular JK-fit auxiliary basis sets is introduced.
The three-center integral tensors and their derivatives required by the density fitting algorithms are often too large to be fully stored in GPU memory.
Consequently, these tensors are divided into small batches, along either the  atomic orbital dimension or the auxiliary basis dimension, and are evaluated in multiple passes.
When computing the derivatives, nested loops over these batches are encountered, which results in some tensors being recomputed and generated multiple times.
This could lead to a relatively high overhead in density fitting TDDFT derivative computations for large molecules. 

The TDDFT-ris implementation is based on the density-fitting variant, except that the auxiliary basis sets are replaced by those specific to TDDFT-ris. The parameters for the J-fit and K-fit basis sets are identical and were optimized to minimize errors in excitation energies.\cite{zhou2023minimal}.
The three-index integral tensor in the TDDFT-ris method is significantly smaller than its counterpart in the density-fitting TDDFT program.
These tensors can be transformed into the molecular orbital representation, retaining only the specific necessary occupied-occupied and occupied-virtual blocks to further reduces their sizes.
Additionally, the molecular orbital space can be truncated to reduce computational and storage costs.
However, in the calculations presented in this work, all molecular orbitals were included.
This reduced-size integral storage scheme is employed during the calculation of excitation energies.

For the calculation of integral derivatives, it is possible to perform a partial transformation for one or two indices of the three-index integral tensors to reduce memory usage and avoid the recomputation of certain intermediates.
However, this alternative scheme would require a distinct code architecture and memory management strategy.
In our current implementation, we did not pursue alternative schemes for the TDDFT-ris method in derivative coupling and gradients computational programs.
This is because the evaluation of integral derivatives is not the computationally dominant step in these calculations, as will be demonstrated in the subsequent section.

The TDDFT amplitudes and molecular orbital coefficients are inherently subject to phase ambiguity, which can lead to uncertain phases in the derivative couplings.
Proper phase handling is essential when tracking the geometric evolution of the system in molecular dynamics simulations.
To ensure phase consistency along a geometric path, we developed a phase-determination function that sets the phase based on the overlap of the excited-state wavefunction with the previous step.
In our implementation, the overlap is computed using the five most dominant contributing determinants.

\section{Benchmark}\label{sec:benchmark}

In this work, the auxiliary basis set employed in the TDDFT-ris method is composed of $s$- and $p$-type orbitals for Coulomb integrals, while only $s$-type orbitals are used for the exchange integrals.
The exponents of the auxiliary Gaussian functions for each atomic type are adopted from Ref.~\citenum{zhou2023minimal}, where they are optimized for the PBE0~\cite{adamo1999toward} XC functional in conjunction with the def2-TZVP~\cite{weigend2005balanced} basis set.
For the density fitting approximation in the remaining components, the def2-universal-JKFIT auxiliary basis~\cite{weigend2008hartree} is utilized.

Although minor discrepancies exist between the results of TDDFT and TDA, this does not influence our comparison of the TDDFT-ris accuracy against that of standard TDDFT calculations.
Similarly, the systematic difference in computational speed between TDA and TDDFT does not impact our performance analysis of the TDDFT-ris method.
Consequently, unless otherwise specified, all calculations are performed using the TDA at the PBE0/def2-TZVP level of theory.

The molecules used for the benchmarks are selected from the benchmark systems included within the GPU4PySCF package.
These molecules and their corresponding system sizes are detailed in Table~\ref{tab:atoms_orbitals}.
All calculations are performed on a single compute node equipped with 6 CPU cores, 30 GB of system memory, and a single NVIDIA Tesla V100 GPU.

\begin{table}[htbp]
  \centering
  \caption{Molecules used in computational performance tests.}
  \label{tab:atoms_orbitals}
  \begin{threeparttable}
    \begin{tabular}{l S[table-format=2] S[table-format=4]}
      \toprule
      {Molecule} & {Number of atoms} & {Number of orbitals} \\
      \midrule
      Vitamin C        & 20 & 420  \\
      Inosine           & 31 & 661  \\
      Bisphenol A      & 33 & 623  \\
      Mg Porphin       & 37 & 848  \\
      Penicillin V     & 42 & 858  \\
      Ochratoxin A     & 45 & 951  \\
      Cetirizine        & 52 & 993  \\
      Tamoxifen         & 57 & 1042 \\
      Raffinose         & 66 & 1246 \\
      Sphingomyelin     & 84 & 1335 \\
      \bottomrule
    \end{tabular}
  \end{threeparttable}
\end{table}

\subsection{Performance Benchmark}\label{subsec:benchmark}

\begin{figure*}[h!tbp]
\subfloat[Exact integrals]{
    \includegraphics[width=0.45\textwidth]{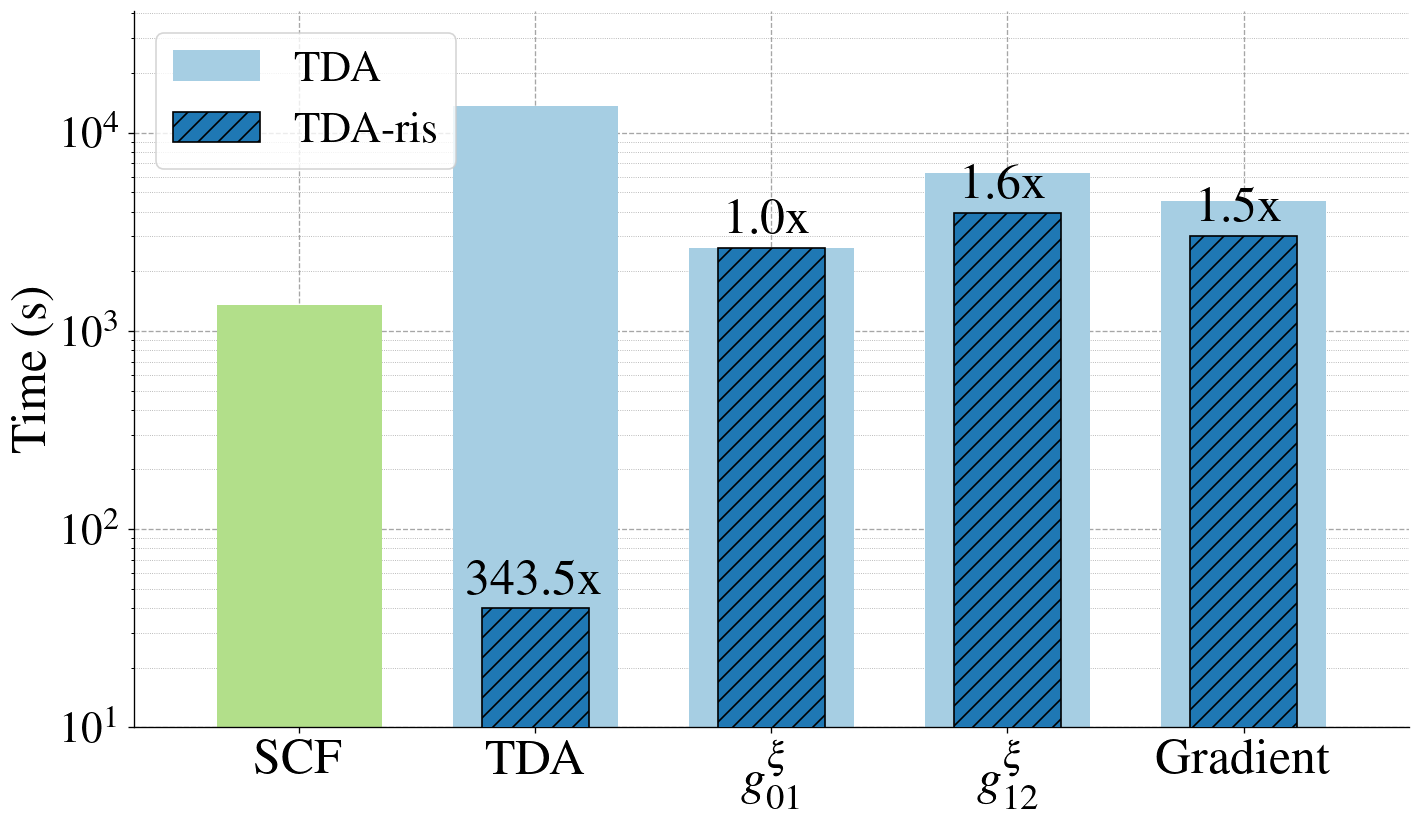} 
    \label{fig:timing_direct}
}   
\subfloat[Density fitting]{
    \includegraphics[width=0.45\textwidth]{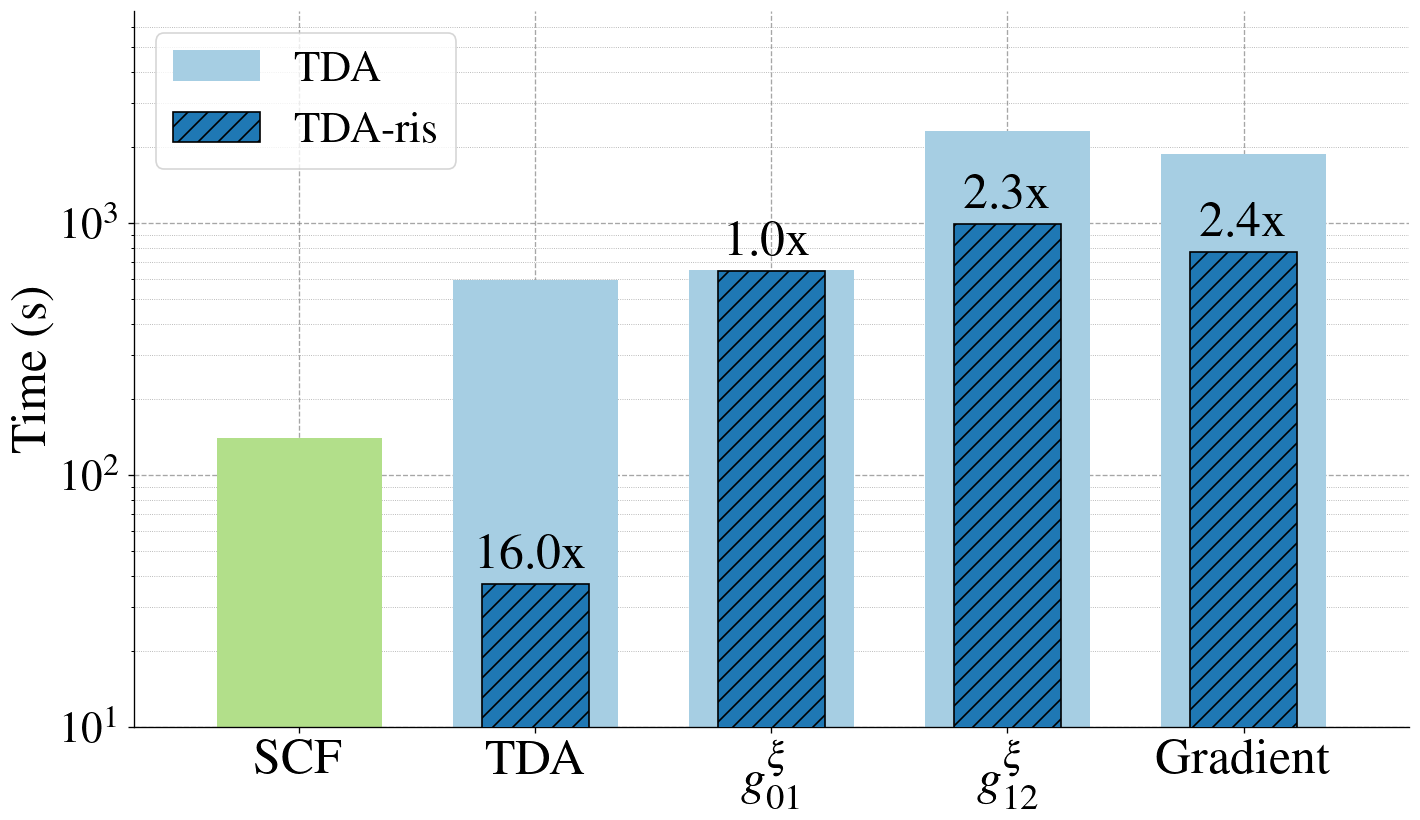} 
    \label{fig:timing_df}
}
\caption{Performance comparison of the TDA-ris and standard TDA methods for calculating the SCF energy, excitation energy, derivative couplings ($\boldsymbol{g}_{01}^{\xi}$ and $\boldsymbol{g}_{12}^{\xi}$), and excited-state gradient based on the total computational time for the molecules listed in Table~\ref{tab:atoms_orbitals}. The two-electron integrals in the ground state and $Z$-vector equations are evaluated with (a) the exact algorithm and (b) density fitting approximation.}
\label{fig:time} 
\end{figure*}

The performance for excited state energy computations, gradients, and derivative couplings calculations using the TDA-ris and TDA approaches is illustrated in Figures~\ref{fig:timing_direct} (exact integrals) and \ref{fig:timing_df} (density fitting).
To clearly demonstrate the cost of the excited state computations, the computational time for the underlying Self-Consistent Field (SCF) calculation is also provided for reference. 

In general, each individual derivative of the excited states consumes 5 to 10 times the cost of the ground state SCF computation.
Owing to the omission of the XC kernel and the aggressive approximation of two-electron integrals, the TDA-ris method offers substantial acceleration for excited-state energy calculations.
Compared to TDA computations with density fitting or exact integrals, the TDA-ris method achieves speed-up factors of 16 and 343, respectively.
Regarding the derivative computations, however, the performance gains of the TDA-ris method are considerably diminished.

Among the derivative computation tasks, no acceleration is observed for the calculation of the ground-to-excited state derivative coupling, $\boldsymbol{g}_{01}^{\xi}$.
A modest speed-up is observed in the benchmark tests for the excited-state-to-excited-state derivative coupling, $\boldsymbol{g}_{12}^{\xi}$, and the excited-state gradient, $\boldsymbol{g}_{1}^{\xi}$.
When the two-electron integrals for the ground state are approximated using regular density fitting (Figure~\ref{fig:timing_df}), the speed-ups of the TDA-ris method are 2.3 and 2.4, respectively.
When exact integrals are employed (Figure~\ref{fig:timing_direct}), the acceleration achieved by the TDA-ris approximation is reduced to 1.6 and 1.5, respectively.

To understand the large performance difference between derivative coupling and excitation energy calculations, we profiled the derivative calculations using both the standard TDDFT and TDDFT-ris implementations.
The computational cost of these calculations arises mainly from three components: solving the $Z$-vector equations, evaluating the exchange–correlation (XC) functional terms, and computing the derivatives of the Coulomb and exchange integrals.
Table~\ref{tab:profiling} presents an example of this profiling for the cetirizine molecule, where the two-electron integrals were evaluated using the density-fitting approximation.
For $\boldsymbol{g}_{01}^{\xi}$, the computational time for each component and the total time are nearly identical between the standard TDA and TDA-ris methods.
This is an expected outcome, as this calculation does not involve any terms that are approximated by the TDA-ris auxiliary integrals.
The generating function, Eq.~\eqref{equ:g0I}, and the resulting Lagrangian, Eq.~\eqref{equ:Lagrangian}, are identical to those within the standard TDA framework; consequently, an identical computational procedure is employed for $\boldsymbol{g}_{01}^{\xi}$ by both TDA and TDA-ris methods.
The speed-up for $\boldsymbol{g}_{12}^{\xi}$ and $\boldsymbol{g}_{1}^{\xi}$ originates from the accelerated evaluation of XC functional terms and the computation of derivatives of Coulomb and exchange integrals.

Standard TDDFT derivative calculations for $\boldsymbol{g}_{12}^{\xi}$ and $\boldsymbol{g}_{1}^{\xi}$ are exceptionally time-consuming, primarily because the requisite third-order derivatives of the XC functionals are currently computed on CPUs.
The transition of the XC computations from CPU to GPU is hampered by unresolved implementation issues related to third-order derivatives.
Once these issues are resolved, an overall performance improvement of approximately a factor of two can be expected.
The TDA-ris method exhibits negligible computational cost for the XC part, as it circumvents the need for third-order derivatives and the computation is completely performed on GPUs.

While the TDA-ris method does provide a speed-up in the calculation of Coulomb and exchange integral derivatives compared to standard TDDFT, the improvement is moderate.
This is because the minimal basis RI approximation in TDA-ris applies to a subset of the terms in this computational step. As shown in Eqs.~\eqref{equ:gI_xi_ris} and \eqref{equ:gIJ_xi_ris}, many terms remain unchanged and do not benefit from the TDA-ris scheme.
Furthermore, when evaluating derivatives of the TDDFT Lagrangian~\eqref{equ:Lagrangian}, it is necessary to compute the ground-state orbital response with respect to nuclear coordinates exactly.
This orbital response is independent of the TDDFT linear-response matrix.
The integral approximation in TDA-ris only reduces the cost associated with the linear-response matrix and the relevant integral derivatives; \textit{i.e.}, solving the $Z$-vector equations does not benefit from the TDA-ris scheme.
As shown in Table~\ref{tab:profiling}, in a typical nuclear gradient calculation, integral derivatives account for roughly 30\% of the computational effort, while solving the $Z$-vector equations for the ground-state orbital response consumes approximately 60\% of the total cost.
Consequently, the computational advantage of the TDA-ris framework diminishes for these higher-order properties.

\begin{figure}[h!tbp]
  \includegraphics[width=0.82\textwidth]{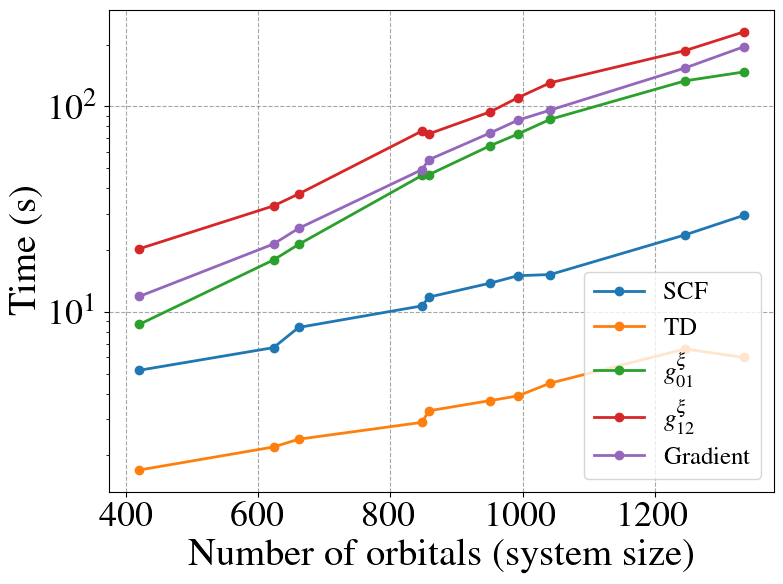} 
  \caption{Scaling of the computational cost for the different stages in a TDA-ris calculation with respect to system size (number of basis functions). The density fitting approximation is employed for the ground state and $Z$-vector computations.}
  \label{fig:timing_scale}
\end{figure}

The scaling of computational time with respect to system size for the TDA-ris excitation energy, derivative coupling, and excited-state gradient calculations is presented in Figure~\ref{fig:timing_scale}.
The evaluations of the derivative coupling and gradient constitute the rate-determining steps of the overall procedure, with computational costs that far exceed those of the SCF and TDA-ris energy calculations.
Considering the requirements of surface-hopping molecular dynamics that employs TDDFT as the electronic structure solver, the dominant expense arises from the computation of excited-state gradients and derivative couplings.
Employing the TDA-ris approximation could result in an overall twofold acceleration.

\begin{table}[htbp]
  \centering
  \caption{Breakdown of computational time (in seconds) for the most computationally intensive tasks in property calculations of TDA and TDA-ris methods. The two-electron integrals in the ground state and $Z$-vector equations are calculated with the density fitting approximation.}
  \label{tab:profiling}
  \begin{tabular}{ll S[table-format=3.1] S[table-format=3.1]}
    \toprule
    {Computational Task} & {Property} & {TDA} & {TDA-ris} \\
    \midrule
    \multirow{3}{*}{Solving $Z$-vector equations} & $\boldsymbol{g}_{01}^{\xi}$ & 48.8 & 46.6 \\
                                             & $\boldsymbol{g}_{12}^{\xi}$ & 61.8 & 62.1 \\
                                             & $\boldsymbol{g}_{1}^{\xi}$  & 54.0 & 54.2 \\
    \midrule
    \multirow{3}{*}{Calculation of XC terms} & $\boldsymbol{g}_{01}^{\xi}$ & 2.2  & 2.3 \\
                                             & $\boldsymbol{g}_{12}^{\xi}$ & 121.2& 2.7 \\
                                             & $\boldsymbol{g}_{1}^{\xi}$  & 119.2& 2.7 \\
    \midrule
    \multirow{3}{*}{\parbox{6cm}{\raggedright Derivatives of Coulomb and exchange integrals}} & $\boldsymbol{g}_{01}^{\xi}$ & 19.9 & 20.3 \\
                                                                   & $\boldsymbol{g}_{12}^{\xi}$ & 70.1 & 36.8 \\
                                                                   & $\boldsymbol{g}_{1}^{\xi}$  & 30.8 & 19.5 \\
    \midrule
    \multirow{3}{*}{Total time}              & $\boldsymbol{g}_{01}^{\xi}$ & 74.8 & 73.5 \\
                                             & $\boldsymbol{g}_{12}^{\xi}$ & 267.5 & 110.5 \\
                                             & $\boldsymbol{g}_{1}^{\xi}$  & 215.4 & 85.3 \\
    \bottomrule
  \end{tabular}
  \end{table}

\subsection{Ris approximation for \textit{Z}-vector Solver}\label{subsec:ris-modified-solver}

Handling the orbital response exactly in derivative calculations limits the performance gains of the TDDFT-ris method.
One might wonder whether the $Z$-vector equations can be accelerated using an approach analogous to TDDFT-ris by applying an approximation to the integrals in the ground-state orbital response.
In this scheme, the approximation in Eq.~\eqref{equ:ris} is applied to the terms within the $(A+B)$ matrix of Eq.~\eqref{equ:Z_vector} during the evaluation of derivative properties.
We test this approximation for the $Z$-vector equations in the calculation of excited-state gradients for the molecules listed in Table~\ref{tab:atoms_orbitals}.
Table~\ref{tab:modified_zvector_solver_timing} summarizes the computational performance and accuracy of this approach compared to the standard TDA-ris method.
The results indicate that the ris-approximated $Z$-vector solver substantially reduces computational time, achieving an acceleration of approximately four- to five-fold for calculations with exact integrals and two- to three-fold when density fitting is employed.
The deviations in the calculated gradients are generally small.

However, approximating the orbital response in this scheme affects only the gradients and derivative couplings, not the excitation energies.
This may disrupt the consistency between energies and derivatives.
Such inconsistencies can slow down the convergence of geometry optimizations or violate energy conservation in molecular dynamics simulations.
Given the significant performance gains, we will investigate in our future work whether applying the ris-approximation to the $Z$-vector solver is suitable for more complex systems and a wider range of properties.

\begin{table}[htbp]
  \centering
  \caption{Comparison of computational time (s) and gradient deviations (Hartree/Bohr) for the TDA-ris (standard) and TDA-ris with ris-approximated $Z$-vector solver (ris solver) methods.}
  \label{tab:modified_zvector_solver_timing}
  \begin{threeparttable}
    \begin{tabular}{l
                    S[table-format=3.1] S[table-format=2.1]
                    S[table-format=3.1] S[table-format=2.1]
                    S[table-format=1.2e-2] S[table-format=1.2e-2]}
      \toprule
      \multirow{2}{*}{Molecule} & \multicolumn{2}{c}{Time (exact integrals)} & \multicolumn{2}{c}{Time (density fitting)} & \multicolumn{2}{c}{Deviation} \\
      \cmidrule(lr){2-3} \cmidrule(lr){4-5} \cmidrule(lr){6-7} 
      & {standard} & {ris solver} & {standard} & {ris solver} & {MAD} & {RMS} \\
      \midrule
      Vitamin C        & 65.7  & 18.5 & 11.9 & 8.3  & 6.36e-04 & 1.03e-03 \\
      Inosine           & 159.7 & 34.3 & 25.5 & 12.9 & 5.15e-04 & 1.64e-03 \\
      Bisphenol A      & 142.5 & 34.9 & 21.4 & 12.4 & 2.63e-04 & 4.41e-04 \\
      Mg Porphin       & 320.8 & 68.2 & 49.3 & 21.2 & 7.37e-05 & 9.86e-05 \\
      Penicillin V     & 324.0 & 62.8 & 55.0 & 21.1 & 2.91e-04 & 5.08e-04 \\
      Ochratoxin A     & 386.0 & 69.7 & 74.3 & 25.5 & 2.96e-04 & 4.88e-04 \\
      Cetirizine        & 447.5 & 88.8 & 85.7 & 29.4 & 5.36e-04 & 9.73e-04 \\
      Tamoxifen         & 375.6 & 78.4 & 96.0 & 31.2 & 1.06e-04 & 1.80e-04 \\
      Raffinose         & 405.7 & 91.9 & 153.9& 45.7 & 3.69e-04 & 8.98e-04 \\
      Sphingomyelin     & 400.5 & 91.4 & 195.2& 49.5 & 3.34e-04 & 1.01e-03 \\
      \bottomrule
    \end{tabular}
  \end{threeparttable}
\end{table}

\section{Applications and discussion}\label{sec:application}

In this section, we evaluate the practical applicability of the derivative computation for the TDA-ris method by testing the errors it introduces in comparison to standard TDDFT methods.
Properties that are frequently utilized in excited-state studies include potential energy surfaces and the interactions of electronic and nuclear motion through non-adiabatic couplings.

To evaluate the reliability of the TDA-ris method for describing the potential energy surface, a series of tests are performed to locate the optimized geometry of the first excited state and to calculate the corresponding emission energy.
To assess the quality of the non-adiabatic couplings, we calculate the derivative couplings, which are the key quantities that govern the transition probabilities between electronic states.
These calculations are performed within the framework of the FSSH methodology\cite{tully1990molecular,tully2012perspective}.
Additionally, the MECPs are computed to evaluate the reliability of the derivative couplings obtained from the TDA-ris method.

\subsection{Excitation State Geometry Optimization and Emission Energy}
\label{sec:pes}

\begin{figure}[htbp]
  \includegraphics[width=0.98\columnwidth]{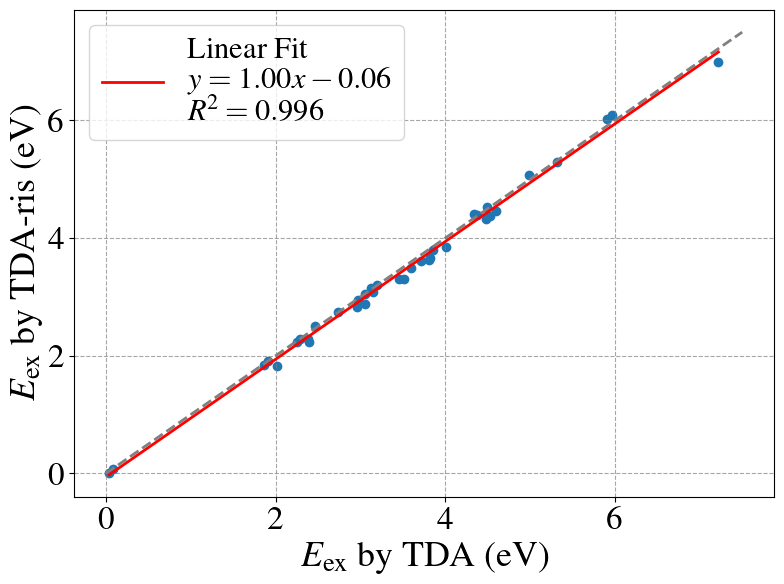}
  \caption{Vertical emission energies (in eV) computed with the TDA-ris method versus the TDA method. For each data point, both the geometry optimization and the energy calculation are performed at the corresponding level of theory. 
  The complete emission energies are presented in Table~\ref{tab:emission}.
  The dashed line indicates perfect agreement ($y=x$), and the solid red line represents the linear regression fit.}
  \label{fig:emission}
\end{figure}

This investigation focuses on the first singlet excited state ($S_1$), irrespective of molecular symmetry.
The test set comprises a diverse range of molecules, with initial geometries taken from Refs.~\citenum{schreiber2008benchmarks} and \citenum{tussupbayev2015comparison}.
The Results are summarized in Table~\ref{tab:emission} and Figure~\ref{fig:emission}.

A close agreement in the emission energies of $S_1$ between the two approaches is observed across the entire set of test molecules, as shown in Figure~\ref{fig:emission}, where TDA-ris emission energies are plotted against their TDA counterparts.
The data points lie closely along the diagonal ($y=x$), indicating minimal systematic deviation and a strong linear relationship.
The emission energies for each molecule from both methods are detailed in Table~\ref{tab:emission}.

To evaluate the quality of the optimized $S_1$ geometries, two metrics are employed.
The first is {\it gradient consistency}, defined as the norm of the TDA gradient calculated at the TDA-ris
optimized geometry and vice versa.
Perfect gradient consistency would yield a zero norm, indicating that the two methods would be able to converge to the same equilibrium geometry under a gradient descent optimization.
As shown in Table~\ref{tab:emission}, these gradient norms are consistently small, on the order of $10^{-3}$~Hartree/Bohr.
This finding indicates that a geometry optimized with one method lies in close proximity to a stationary point on the $S_1$ potential energy surface of the other method.

The second metric involves direct structural comparisons between the optimized geometries, quantified by the mean absolute difference (MAD) and root-mean-square (RMS) difference.
For all tested molecules, these deviations are small, demonstrating that TDA-ris produces optimized first excited-state structures highly consistent with those from the standard TDA method.

In summary, the results indicate that the TDA-ris method yields emission energies and optimized geometries that are in strong agreement with those obtained from the standard TDA method.
This level of agreement substantiates the use of TDA-ris as a reliable and efficient method for computationally demanding tasks, such as geometry optimizations in fluorescence studies.
Nevertheless, a degree of caution is warranted, as we occasionally observe that the TDA-ris approximation can alter the energetic ordering of excited states.
Consequently, it is crucial to verify that the state of interest in a TDA-ris calculation corresponds to the correct electronic state within the standard TDA framework, particularly in systems with a high density of states.
Despite this minor limitation, the performance benefits justify the application of the method.

\begin{table*}[htbp]
\centering
\caption{Comparison of TDA and TDA-ris for excited-state emission energies and excited-state geometries.}
\label{tab:emission}
\begin{threeparttable}
  \begin{tabular}{p{16ex}
		  S[table-format=1.2]
		  S[table-format=1.2]
		  S[table-format=1.2e-2]
		  S[table-format=1.2e-2]
		  S[table-format=1.2e-2]
		  S[table-format=1.2e-2]}
    \toprule
\multirow{2}{*}{Molecule} & \multicolumn{2}{c}{Emission energy (eV)} &
\multicolumn{2}{c}{Gradients consistency\tnote{a}} &
\multicolumn{2}{c}{Geometry consistency\tnote{b}} \\
\cmidrule(lr){2-3} \cmidrule(lr){4-7}
& {TDA} & {TDA-ris} & {TDA-ris} & {TDA} & {Geom. MAD} & {Geom. RMS} \\
    \midrule
      acetamide       & 4.53 & 4.38 & 2.89e-03 & 2.83e-03 & 1.79e-03 & 2.86e-03 \\
      acetone         & 3.81 & 3.62 & 1.29e-03 & 1.36e-03 & 2.08e-03 & 3.18e-03 \\
      adenine         & 3.85 & 3.79 & 4.10e-03 & 4.11e-03 & 5.94e-04 & 8.80e-04 \\
      benzene         & 5.32 & 5.29 & 2.31e-03 & 2.32e-03 & 3.54e-04 & 4.95e-04 \\
      benzoquinone    & 2.39 & 2.23 & 2.40e-03 & 1.91e-03 & 2.70e-04 & 6.34e-04 \\
      butadiene       & 5.91 & 6.03 & 1.03e-02 & 1.03e-02 & 2.52e-03 & 3.71e-03 \\
      cyclopentadiene & 4.49 & 4.53 & 6.87e-03 & 6.95e-03 & 5.63e-04 & 9.22e-04 \\
      cytosine        & 2.97 & 3.71 & 1.51e-03 & 4.95e-03 & 2.27e-02 & 4.30e-02 \\
      ethene          & 7.22 & 6.99 & 1.98e-03 & 1.99e-03 & 6.18e-04 & 8.35e-04 \\
      formaldehyde    & 3.51 & 3.31 & 1.87e-03 & 1.89e-03 & 5.43e-04 & 9.23e-04 \\
      formamide       & 4.48 & 4.33 & 2.86e-03 & 2.82e-03 & 1.15e-03 & 1.59e-03 \\
      furan           & 5.97 & 6.10 & 7.53e-03 & 7.56e-03 & 5.10e-04 & 7.76e-04 \\
      hexatriene      & 4.99 & 5.07 & 7.25e-03 & 7.19e-03 & 1.41e-03 & 2.25e-03 \\
      imidazole       & 0.04 & 0.01 & 4.24e-02 & 9.81e-02 & 1.03e-02 & 1.58e-02 \\
      naphthalene     & 4.38 & 4.40 & 1.44e-03 & 1.51e-03 & 3.67e-04 & 5.30e-04 \\
      norbornadiene   & 2.38 & 2.28 & 4.70e-03 & 4.94e-03 & 2.51e-03 & 3.55e-03 \\
      octatetraene    & 4.34 & 4.41 & 5.19e-03 & 5.14e-03 & 1.16e-03 & 1.70e-03 \\
      propanamide     & 4.60 & 4.46 & 3.00e-03 & 2.98e-03 & 1.85e-03 & 2.88e-03 \\
      pyrazine        & 3.82 & 3.66 & 1.94e-03 & 1.94e-03 & 2.59e-04 & 4.69e-04 \\
      pyridazine      & 3.05 & 2.88 & 8.42e-03 & 8.01e-03 & 1.10e-03 & 1.75e-03 \\
      pyridine        & 4.01 & 3.85 & 3.66e-03 & 3.63e-03 & 1.36e-03 & 2.01e-03 \\
      pyrimidine      & 3.46 & 3.31 & 2.11e-03 & 2.11e-03 & 3.27e-04 & 5.18e-04 \\
      pyrrole         & 0.08 & 0.07 & 8.90e-04 & 1.92e-03 & 4.89e-02 & 6.83e-02 \\
      tetrazine       & 2.02 & 1.83 & 1.06e-03 & 1.12e-03 & 1.48e-04 & 2.16e-04 \\
      thymine         & 3.71 & 3.61 & 4.74e-03 & 4.68e-03 & 9.07e-04 & 1.55e-03 \\
      triazine        & 2.97 & 2.91 & 2.85e-03 & 3.12e-03 & 3.28e-02 & 4.80e-02 \\
      uracil          & 3.60 & 3.50 & 4.76e-03 & 5.29e-03 & 2.74e-03 & 4.40e-03 \\
      coronene        & 3.20 & 3.21 & 1.28e-03 & 1.37e-03 & 7.70e-05 & 1.15e-04 \\
      coumarin 153    & 3.05 & 3.05 & 1.50e-03 & 1.49e-03 & 1.13e-01 & 1.45e-01 \\
      m1$^a$ & 3.15 & 3.09 & 3.63e-03 & 4.00e-03 & 8.05e-04 & 1.02e-03 \\
      Indigo          & 1.86 & 1.85 & 1.08e-03 & 9.70e-04 & 1.02e-03 & 1.67e-03 \\
      naphtacene      & 2.47 & 2.50 & 1.61e-03 & 1.69e-03 & 7.20e-05 & 1.18e-04 \\
      phtalocyanine   & 2.25 & 2.24 & 1.36e-03 & 1.46e-03 & 1.08e-04 & 1.67e-04 \\
      m2$^b$ & 2.74 & 2.74 & 1.69e-03 & 1.70e-03 & 2.28e-02 & 2.98e-02 \\
    \bottomrule
    \end{tabular}
    \begin{tablenotes}
      \small
      \item[a] Norm of excited-state gradients evaluated at geometries optimized by the other method (unit in Hartree/Bohr).
      \item[b] Mean absolute difference (MAD) and root-mean-square (RMS) difference between geometries by TDA and TDA-ris (unit in Bohr).
    \end{tablenotes}
  \end{threeparttable}
\end{table*}

\subsection{derivative coupling}
\label{sec:derivative:coupling}

\begin{figure}[h!tbp]
  \centering
  
  \subfloat[Coupling between the $S_0$ and $S_1$ states.]{
    \includegraphics[width=0.47\columnwidth]{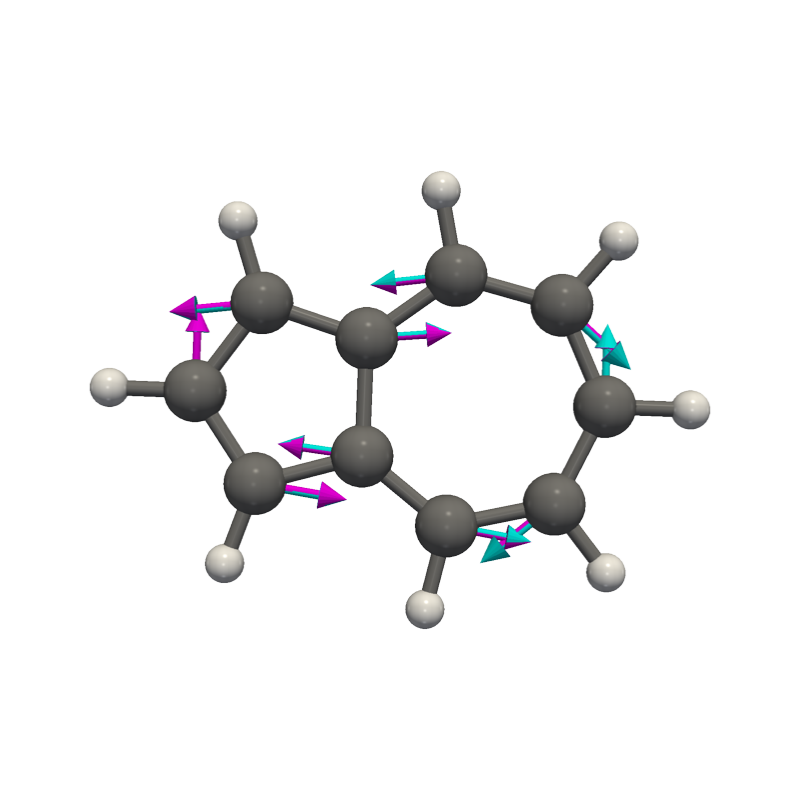}
    \label{fig:azulene_ge}
  }
  \hfill 
  \subfloat[Coupling between the $S_1$ and $S_2$ states.]{
    \includegraphics[width=0.47\columnwidth]{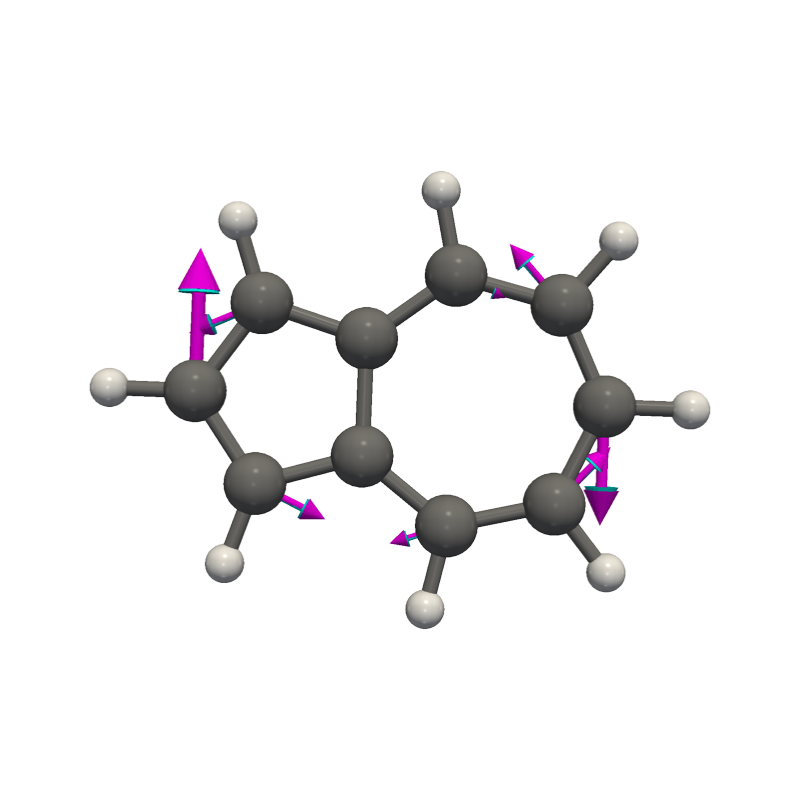}
    \label{fig:azulene_ee}
  }
  
  \caption{Derivative couplings for azulene. The vectors derived from the TDA and TDA-ris methods are depicted as magenta and cyan arrows, respectively, illustrating the coupling between the (a) $S_0/S_1$ ($\boldsymbol{g}_{01}^{\xi}$) and (b) $S_1/S_2$ ($\boldsymbol{g}_{12}^{\xi}$) electronic states.}
  \label{fig:azulene_nacv}
\end{figure}

Derivative couplings are fundamental quantities for modeling non-adiabatic molecular dynamics\cite{prlj2025best,jain2022pedagogical,curchod2018ab}.
The test set includes a subset of molecules with geometries adopted from Ref.~\citenum{schreiber2008benchmarks}.
Table~\ref{tab:app_fonacme} summarizes the MAD and RMS deviation of derivative couplings computed using the TDA-ris method against reference TDA values.
The analysis encompasses couplings between the ground state ($S_0$) and the first excited state ($S_1$), denoted as $\boldsymbol{g}_{01}^{\xi}$, and between the $S_1$ and the second excited state ($S_2$), denoted as $\boldsymbol{g}_{12}^{\xi}$.
As an illustration, Figure~\ref{fig:azulene_nacv} depicts the derivative coupling vectors for azulene at its ground-state equilibrium geometry.
Specifically, Figure~\ref{fig:azulene_ge} compares the $\boldsymbol{g}_{01}^{\xi}$ vectors, while Figure~\ref{fig:azulene_ee} presents the corresponding comparison for the $\boldsymbol{g}_{12}^{\xi}$ vectors.

The results in Table~\ref{tab:app_fonacme} indicate that the $\boldsymbol{g}_{01}^{\xi}$ couplings calculated with TDA-ris exhibit excellent agreement with the reference TDA values.
In contrast, the agreement for couplings between excited states, $\boldsymbol{g}_{12}^{\xi}$, is significantly less accurate.
This trend is illustrated for azulene in Figure~\ref{fig:azulene_nacv}.
The $\boldsymbol{g}_{01}^{\xi}$ vectors computed by both methods are nearly identical.
In contrast, the $\boldsymbol{g}_{12}^{\xi}$ vectors are aligned in direction but show a larger difference in magnitude, indicating larger deviations in the TDA-ris wavefunctions for the higher excited states.

For certain molecules, significant discrepancies are observed in the derivative couplings $\boldsymbol{g}_{12}^{\xi}$. 
Cytosine, for example, exhibits a significant deviation.
The MAD and RMS deviations for $\boldsymbol{g}_{12}^{\xi}$ between the TDA and TDA-ris methods are $21.7$ and $50.1$~$a_0^{-1}$, respectively.
This deviation can be attributed to the discrepancy in the predicted energies of the $S_1$ and $S_2$ states between the two methods.
The reference TDA calculation predicts a near-degeneracy between the $S_1$ and $S_2$ states, with energies of $4.930$~eV and $4.932$~eV, respectively.
The TDA-ris method, in contrast, predicts a larger energy gap, with corresponding energies of $4.865$~eV and $4.896$~eV.
In the expression for the derivative coupling, the energy difference between the coupled electronic states appears in the denominator.
Consequently, the small denominator arising from the small energy gap in the reference TDA calculation results in a substantially larger derivative coupling magnitude compared to that from the TDA-ris method.
For comparison, the MAD and RMS deviations for the numerator term (the non-adiabatic coupling vector) of the derivative couplings are merely $1.2 \times 10^{-4}$ and $3.2 \times 10^{-4}$~$\text{Hartree}/a_0^{-1}$, respectively.
Similar large discrepancies for molecules such as cyclopentadiene, ethene, furan, and triazine also arise from this same issue.

Furthermore, the electronic structure of an excited state, as characterized by its transition vector, may differ between the TDA-ris and TDA methods.
This discrepancy leads to differences in the numerator of the derivative coupling vector.
When such differences in the numerator (originating from the transition vector) are compounded by small energy gaps in the denominator (due to near-degeneracy of electronic states), significant deviations between the results of the two methods will be anticipated.
These issues are also manifested in the calculation of the hopping probabilities, as detailed in the subsequent section.
Consequently, scenarios involving near-degenerate states, state reordering, or substantial disparities in the excited-state electronic structures between the TDA and TDA-ris formalisms must be treated with caution.
Such conditions can severely compromise the accuracy of derivative coupling calculated using the TDA-ris method.

\begin{table*}[htbp]
  \centering
  \caption{Comparison of derivative couplings for $S_0-S_1$ and $S_1-S_2$ using TDA and TDA-ris methods (unit in $ a_0^{-1} $).}
  \label{tab:app_fonacme}
  \begin{threeparttable}
    \begin{tabular}{l
                    S[table-format=1.2e-2]
                    S[table-format=1.2e-2]
                    S[table-format=1.2e+1]
                    S[table-format=1.2e+1]}
      \toprule
      \multirow{2}{*}{Molecule} & \multicolumn{2}{c}{S$_0$-S$_1$} & \multicolumn{2}{c}{S$_1$-S$_2$} \\
      \cmidrule(lr){2-3} \cmidrule(lr){4-5} 
      & {MAD} & {RMS} & {MAD} & {RMS} \\
      \midrule
      acetamide       & 8.74e-04 & 1.58e-03 & 9.73e-03 & 1.60e-02 \\
      acetone         & 5.31e-04 & 8.13e-04 & 2.74e-03 & 4.06e-03 \\
      adenine         & 6.23e-04 & 1.35e-03 & 2.36e-02 & 6.26e-02 \\
      azulene         & 7.06e-04 & 1.23e-03 & 4.57e-03 & 8.60e-03 \\
      benzene         & 1.57e-03 & 2.38e-03 & 6.85e-03 & 9.93e-03 \\
      benzoquinone    & 3.86e-04 & 1.08e-03 & 5.22e-02 & 9.76e-02 \\
      butadiene       & 2.20e-03 & 3.63e-03 & 2.17e-02 & 3.63e-02 \\
      cyclopentadiene & 1.45e-03 & 2.61e-03 & 5.58e-03 & 9.97e-03 \\
      cyclopropene    & 1.14e-03 & 2.55e-03 & 5.21e-01 & 1.05e+00 \\
      cytosine        & 9.57e-04 & 1.47e-03 & 2.17e+01 & 5.01e+01 \\
      ethene          & 3.27e-04 & 6.94e-04 & 3.42e-01 & 7.25e-01 \\
      formaldehyde    & 6.50e-04 & 1.59e-03 & 8.11e-03 & 1.67e-02 \\
      formamide       & 8.75e-04 & 2.01e-03 & 6.09e-03 & 1.43e-02 \\
      furan           & 1.51e-03 & 2.74e-03 & 1.60e+00 & 3.13e+00 \\
      hexatriene      & 1.52e-03 & 2.59e-03 & 5.57e-03 & 9.53e-03 \\
      imidazole       & 1.08e-03 & 2.17e-03 & 4.63e-02 & 7.85e-02 \\
      naphthalene     & 6.82e-04 & 9.28e-04 & 8.97e-02 & 1.78e-01 \\
      norbornadiene   & 9.25e-04 & 1.90e-03 & 2.69e-02 & 4.20e-02 \\
      octatetraene    & 1.11e-03 & 1.73e-03 & 2.93e-03 & 4.97e-03 \\
      propanamide     & 6.94e-04 & 1.35e-03 & 7.32e-03 & 1.36e-02 \\
      pyrazine        & 6.58e-04 & 1.16e-03 & 7.23e-02 & 1.39e-01 \\
      pyridazine      & 9.99e-04 & 2.07e-03 & 9.53e-02 & 1.95e-01 \\
      pyridine        & 6.48e-04 & 1.39e-03 & 2.02e-01 & 4.15e-01 \\
      pyrimidine      & 1.10e-03 & 2.12e-03 & 1.45e-01 & 2.99e-01 \\
      pyrrole         & 8.13e-04 & 2.17e-03 & 6.01e-03 & 1.26e-02 \\
      tetrazine       & 4.78e-04 & 1.28e-03 & 4.24e-02 & 7.97e-02 \\
      thymine         & 3.38e-04 & 6.90e-04 & 8.16e-03 & 1.66e-02 \\
      triazine        & 3.30e-05 & 8.14e-05 & 6.03e+00 & 1.06e+01 \\
      uracil          & 5.16e-04 & 1.11e-03 & 4.00e-03 & 8.97e-03 \\
      \bottomrule
    \end{tabular}
  \end{threeparttable}
\end{table*}

\subsection{Minimum-Energy Crossing Point}

A Minimum-Energy Crossing Point (MECP) is the molecular configuration of lowest energy on the crossing seam where the potential energy surfaces of two or more electronic states are degenerate\cite{prlj2025best,Zhang2025}.
In the vicinity of an MECP, the electronic degeneracy is lifted by infinitesimal geometric distortions within the branching plane.
This plane is determined by two vectors: the derivative coupling vector and the gradient-difference vector ($\boldsymbol{x_1} = \boldsymbol{g}^{\xi}_J - \boldsymbol{g}^{\xi}_I$).
Locating an MECP requires optimization algorithms that utilize both of these vectors.
The search for an MECP resembles the geometry optimization of an excited state, with the distinction that it necessitates the evaluation of derivative coupling vectors in addition to nuclear gradients.
Following the procedure detailed in Ref.~\citenum{bearpark1994direct}, the effective gradient $ \boldsymbol{\widetilde{g}} $ is defined as
\begin{align}
  \boldsymbol{\widetilde{g}} = & \boldsymbol{f} + \boldsymbol{g},
\end{align}
where $\boldsymbol{f} = (E_J - E_I)\frac{\boldsymbol{x_1}}{|\boldsymbol{x_1}|}$ is the gradient component that minimizes the energy difference between the $S_J$ and $S_I$ states. The second term, $\boldsymbol{g} = P \boldsymbol{g}^{\xi}_{J}$, is the gradient of the $S_J$ state, projected by the operator $P$ onto the orthogonal complement of the plane spanned by $\boldsymbol{x_1}$ and $\boldsymbol{g}^{\xi}_{IJ}$, which serves to locate the lowest energy point on the crossing seam.
The resulting effective gradients are then supplied to geometry optimization routines to locate the MECP.

\begin{figure}[h!tbp] 
  \centering 
  \includegraphics[width=0.23\textwidth]{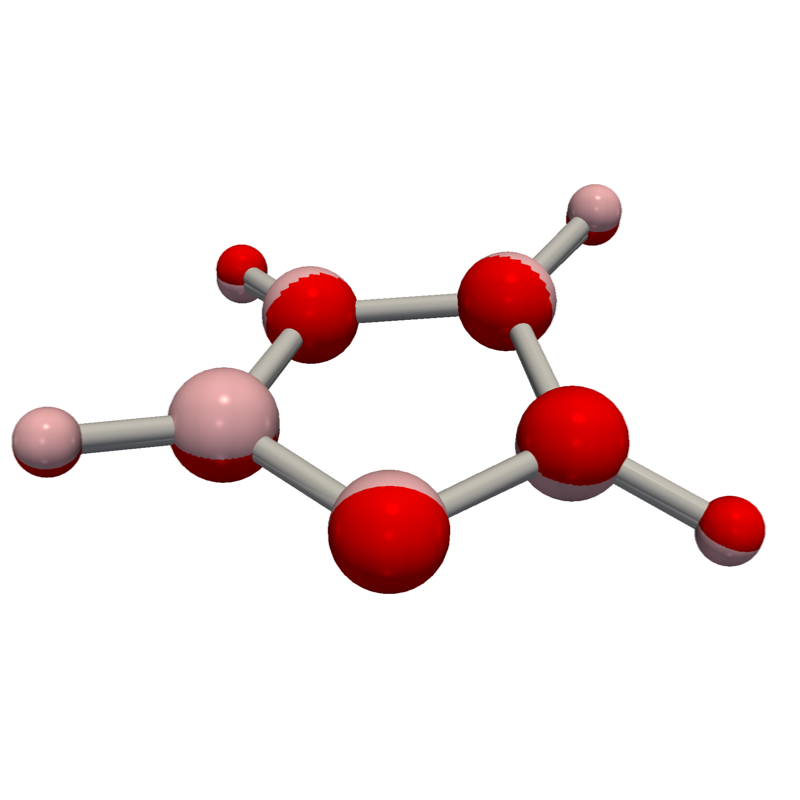}%
  \caption{The geometry of the $S_1/S_2$ minimum energy crossing point of furan using TDA (pink atoms) and TDA-ris (red atoms).}
  \label{fig:mecp_geometries}
\end{figure}

We use the furan molecule as a test system and apply both the TDA and TDA-ris methods to locate the MECP between the $S_1$ and $S_2$ states.
Figure~\ref{fig:mecp_geometries} compares the MECP geometries optimized using the two methods.
The geometries obtained from TDA (pink) and TDA-ris (red) are highly similar.
The TDA-ris method yields an MECP geometry that is structurally analogous to that obtained from the standard TDA method, exhibiting only minor discrepancies.

\begin{figure}[h!tbp] 
  \centering 

  \subfloat[TDA potential energy surface in the branching plane\label{fig:tda_sub}]{%
      \includegraphics[width=0.5\textwidth]{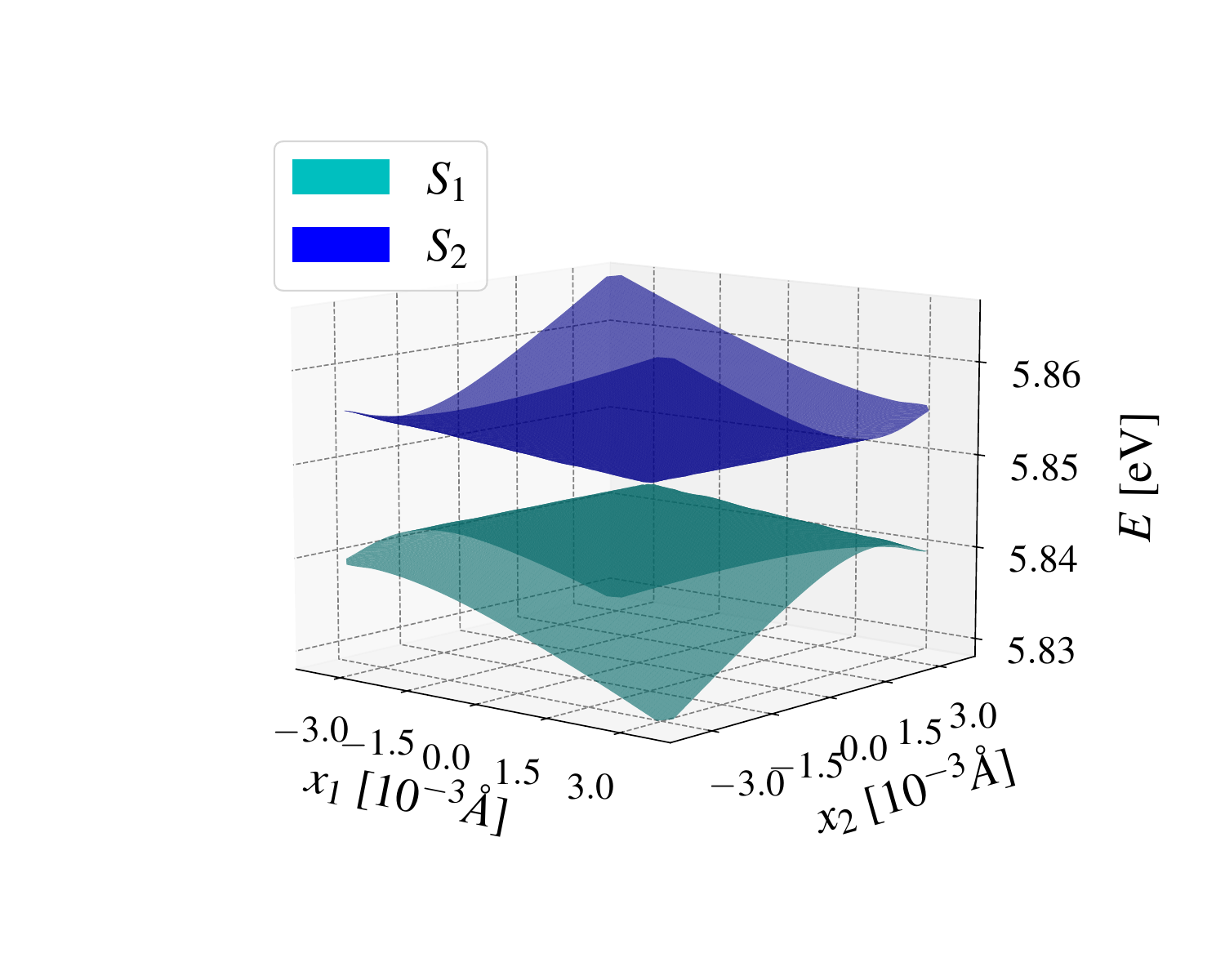}%
  }
  \hfill 
  \subfloat[TDA-ris potential energy surface in the branching plane\label{fig:ris_sub}]{%
      \includegraphics[width=0.5\textwidth]{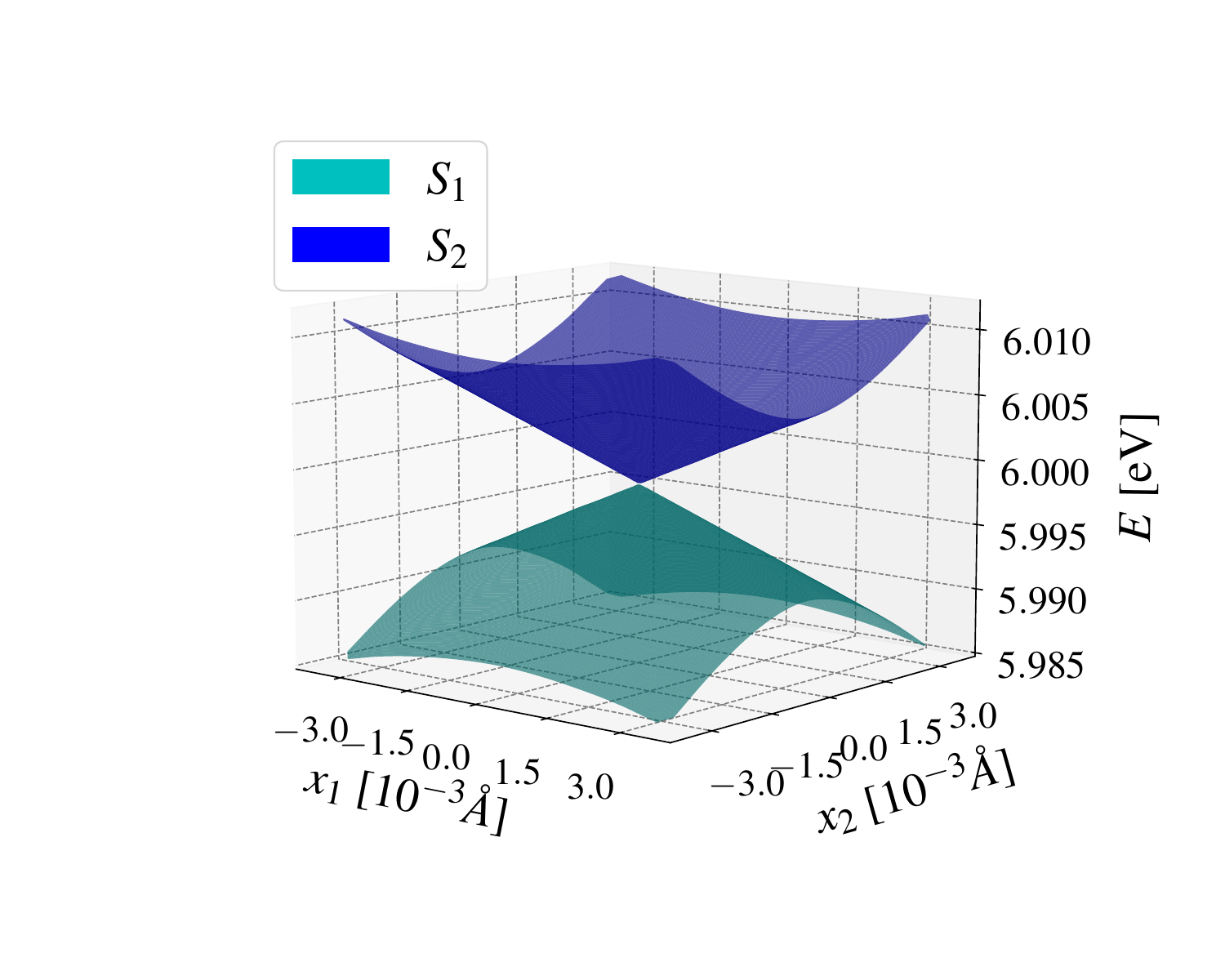}%
  }
  
  \caption{Potential energy surfaces within the branching plane of the $S_1/S_2$ minimum energy crossing point of furan, determined using (a) the TDA and (b) the TDA-ris methods, respectively.
}
  \label{fig:furan_combined}
\end{figure}

Using these optimized geometries, the potential energy surfaces of the $S_1$ and $S_2$ states are plotted within their respective branching planes, as shown in Figure~\ref{fig:tda_sub} (TDA) and Figure~\ref{fig:ris_sub} (TDA-ris).
The overall topology of the potential energy surfaces near the MECP is comparable for both methods.
Subtle differences in structural details are observed.
A primary distinction lies in the absolute energy of the crossing seam. 
The TDA-ris method (Figure~\ref{fig:ris_sub}) places the MECP at approximately $5.998$ eV, representing a significant upward energy shift compared to the $5.847$ eV predicted by the standard TDA calculation (Figure~\ref{fig:tda_sub}). 
Furthermore, the local topographies of the potential energy surfaces differ in their symmetry. 
The standard TDA result exhibits a pronounced anisotropy within the branching plane; the surface is relatively flat along one diagonal, with an energy difference of only $0.015$ eV between its endpoints, but exhibits a much steeper gradient along the orthogonal diagonal, with an energy difference of $0.036$ eV. 
In contrast, the TDA-ris method yields a more isotropic potential energy surface, where the energy variation is more uniform across the surface, resulting in a comparable energy difference of approximately $0.025$ eV along both diagonals.

Despite these discrepancies in absolute energy and surface topography, the fundamental geometric structure of the MECP is qualitatively conserved between the two methods. 
Both calculations robustly predict a well-defined conical intersection, which is the essential feature governing non-adiabatic transitions at this point of degeneracy. 
This suggests that while the finer quantitative details of the potential energy landscape are sensitive to the chosen computational approach, the core physical feature of the $S_1$-$S_2$ state crossing is reliably captured by both TDA and TDA-ris.
This finding, combined with the results for emission energies (Figure~\ref{fig:emission}), confirms the suitability of the TDA-ris method for describing potential energy surface properties, including the equilibrium geometries of excited states and their intersections.

\subsection{Hopping Possibilities in Fewest Switches Surface Hopping Simulations}

\begin{figure}[h!tbp]
    \centering
    \subfloat[Formaldehyde $S_0/S_1$]{
        \label{fig:fssh_formaldehyde_ge}
        \includegraphics[width=0.48\columnwidth]{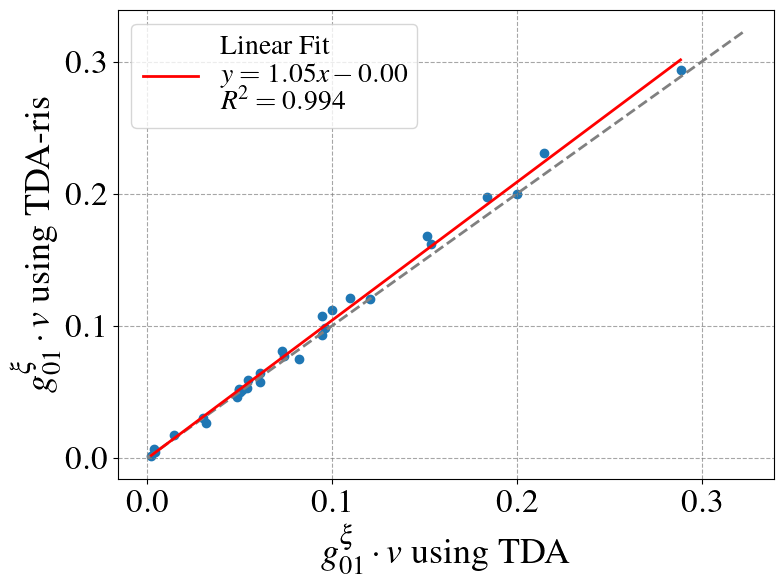}
    }
    \subfloat[Formaldehyde $S_1/S_2$]{
        \label{fig:fssh_formaldehyde_ee}
        \includegraphics[width=0.48\columnwidth]{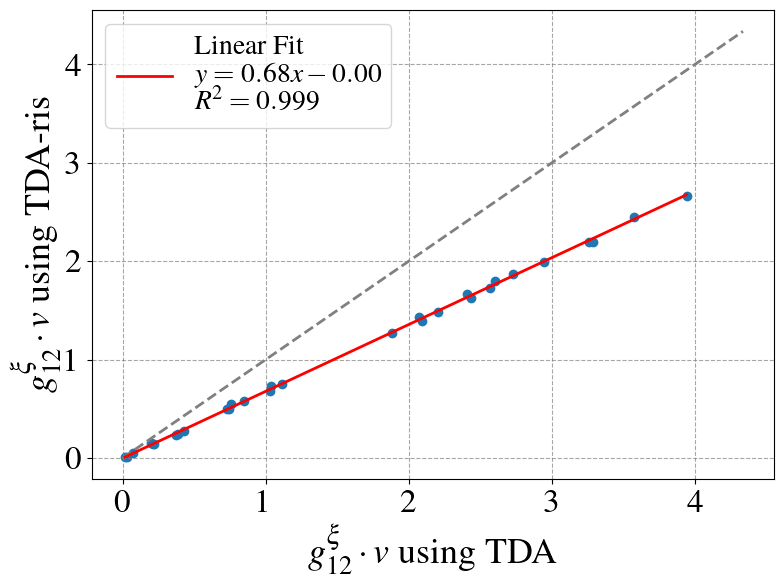}
    }
    \caption{Correlation of time-derivative non-adiabatic coupling matrix elements ($T_{IJ}=\boldsymbol{g}_{IJ}^{\xi} \cdot \boldsymbol{v}$) from TDA and TDA-ris calculations for formaldehyde.}
\label{fig:fssh_formaldehyde}
\end{figure}

\begin{figure}[h!tbp]
    \centering
    \subfloat[Ethene $S_0/S_1$]{
        \label{fig:fssh_ethene_ge}
        \includegraphics[width=0.48\columnwidth]{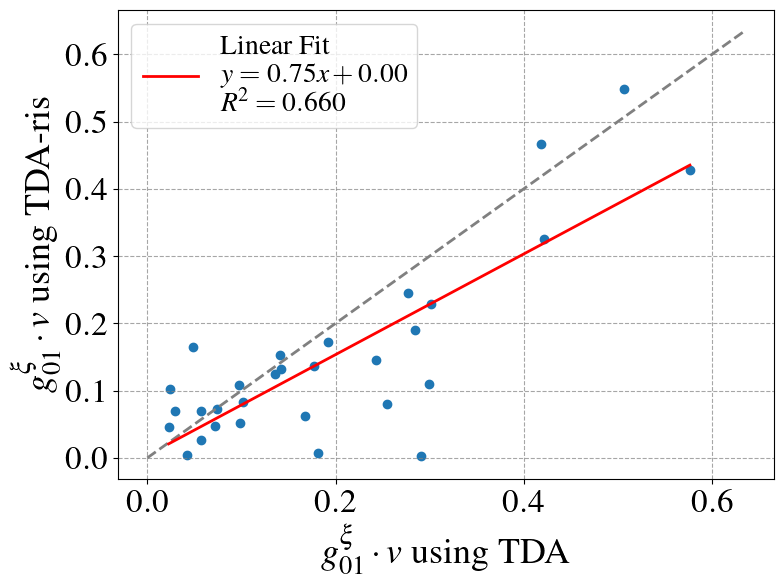}
    }
    \subfloat[Ethene $S_1/S_2$]{
        \label{fig:fssh_ethene_ee}
        \includegraphics[width=0.48\columnwidth]{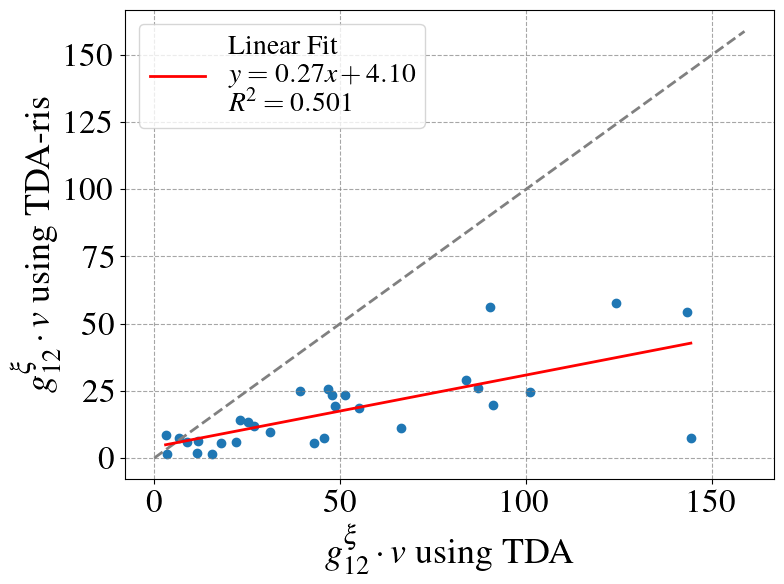}
    }
    \caption{Correlation of time-derivative non-adiabatic coupling matrix elements from TDA and TDA-ris calculations for ethene.}
\label{fig:fssh_ethene}
\end{figure}

\begin{figure}[h!tbp]
    \centering
    \subfloat[Acetone $S_0/S_1$]{
        \label{fig:fssh_acetone_ge}
        \includegraphics[width=0.48\columnwidth]{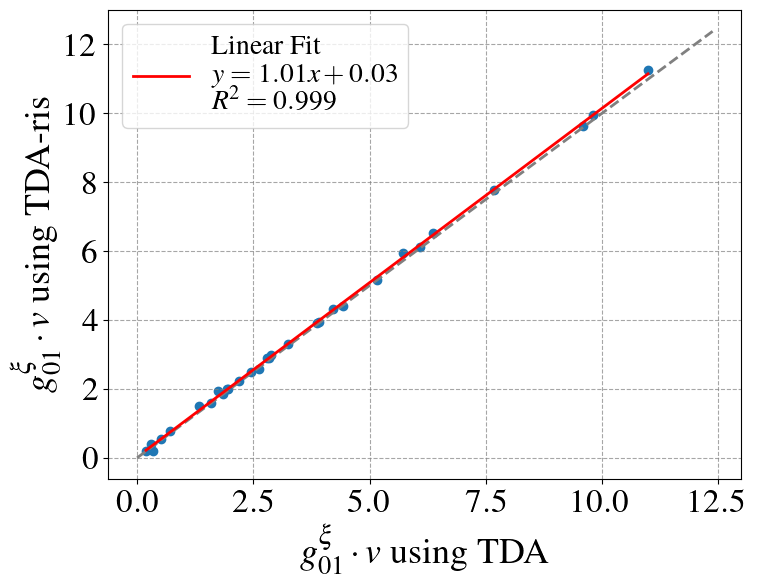}
    }
    \subfloat[Acetone $S_1/S_2$]{
        \label{fig:fssh_acetone_ee}
        \includegraphics[width=0.48\columnwidth]{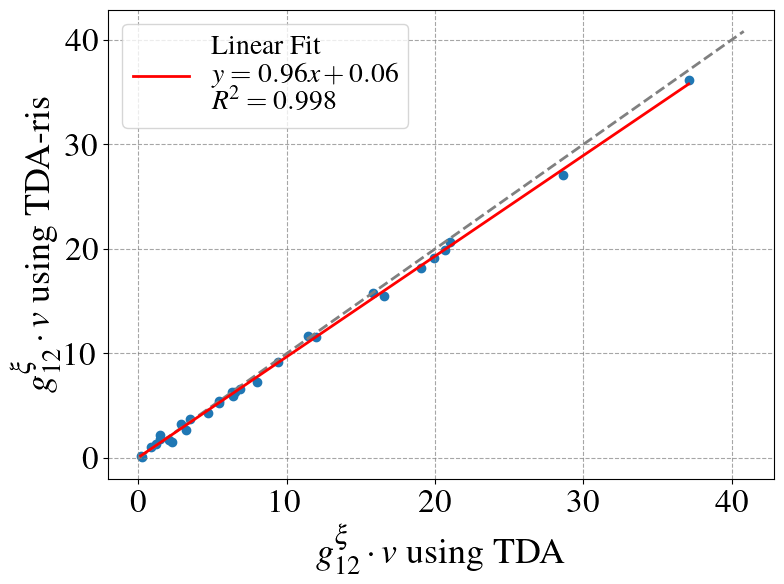}
    }
    \caption{Correlation of time-derivative non-adiabatic coupling matrix elements from TDA and TDA-ris calculations for acetone.}
\label{fig:fssh_acetone}
\end{figure}

\begin{figure}[h!tbp]
    \centering
    \subfloat[Hexatriene $S_0/S_1$]{
        \label{fig:fssh_hexatriene_ge}
        \includegraphics[width=0.48\columnwidth]{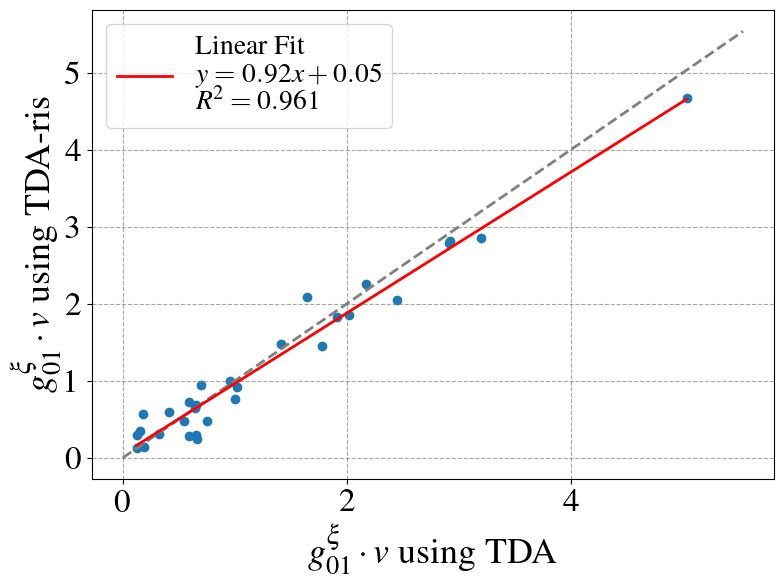}
    }
    \subfloat[Hexatriene $S_1/S_2$]{
        \label{fig:fssh_hexatriene_ee}
        \includegraphics[width=0.48\columnwidth]{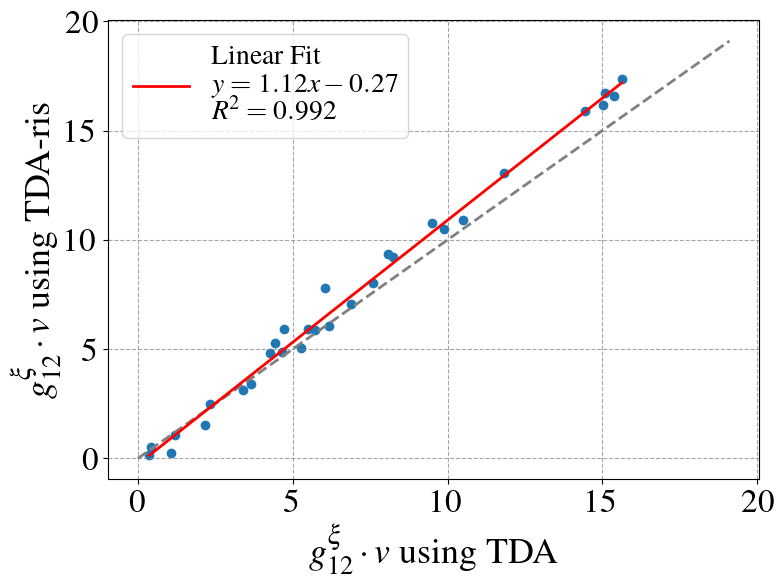}
    }
    \caption{Correlation of time-derivative non-adiabatic coupling matrix elements from TDA and TDA-ris calculations for hexatriene.}
\label{fig:fssh_hexatriene}
\end{figure}

Another application of non-adiabatic couplings is the calculation of hopping probabilities in FSSH algorithm.
To evaluate the applicability of the TDA-ris method in this context, we examine the time-derivative non-adiabatic coupling matrix element, $T_{IJ}$.
This term is proportional to the FSSH transition probability and is defined as $T_{IJ} = \boldsymbol{g}_{IJ}^{\xi} \cdot \boldsymbol{v}$, where $\boldsymbol{v}$ denotes the nuclear velocity vector \cite{jain2022pedagogical}.
Consequently, the $T_{IJ}$ values computed using the TDA-ris method are compared with those obtained from the reference TDA method.
Formaldehyde, ethene, acetone, and hexatriene are selected as test molecules.
For each molecule, a set of initial conditions (geometries and velocities) is generated via Wigner sampling~\cite{peng2019treatment,wigner1932quantum} to compute the $T_{IJ}$ values. 
Based on initial structures from Ref. \citenum{schreiber2008benchmarks}, 30 geometries are sampled at 300 K and are provided in the Supporting Information.
The correlation between the results from the two methods is illustrated in Figures~\ref{fig:fssh_formaldehyde} to \ref{fig:fssh_hexatriene}.

The results reveal significant performance variations contingent upon the coupled electronic states and molecules under consideration.
For the derivative coupling between the ground state ($S_0$) and the first excited state ($S_1$), denoted as $\boldsymbol{g}_{01}^{\xi}$, and that between the first ($S_1$) and second ($S_2$) excited states, $\boldsymbol{g}_{12}^{\xi}$, the TDA and TDA-ris methods generally exhibit a strong linear correlation.
However, the $\boldsymbol{g}_{01}^{\xi}$ coupling in ethene (Figures~\ref{fig:fssh_ethene_ge}) and the $\boldsymbol{g}_{12}^{\xi}$ coupling in both formaldehyde and ethene (Figure~\ref{fig:fssh_formaldehyde_ee} and \ref{fig:fssh_ethene_ee}) are notable exceptions to this trend.
The correlation line deviates significantly from the ideal $y=x$ line.

The deviations can be attributed to two factors.
First, for certain systems, there are significant systematic discrepancies in the excitation energy gaps predicted by the TDA and TDA-ris methods.
This discrepancy introduces a scaling factor in the denominator of the derivative coupling expression, which amplifies the error in the coupling vectors, as discussed in Section~\ref{sec:derivative:coupling}.
The $\boldsymbol{g}_{12}^{\xi}$ couplings for formaldehyde and ethene fall into this category.
This factor is directly responsible for the deviation of the correlation slope from 1.0, as shown in Figures~\ref{fig:fssh_formaldehyde_ee} and \ref{fig:fssh_ethene_ee}.

A second cause for the particularly poor linear correlation observed for ethene stems from a more fundamental issue.
This discrepancy originates from an intrinsic limitation of the TDA-ris approach, arising from the differences in the excited-state electronic structures (i.e., transition vectors) computed by the TDA and TDA-ris methods at certain molecular geometries.
For instance, across the sampled geometries of ethene, the maximum discrepancies in the components of $\boldsymbol{g}_{01}^{\xi}$ between the two methods span a wide range, from $0.0017$ to $0.0200$~$a_0^{-1}$.
In contrast, this range is considerably narrower for hexatriene, spanning from $0.0093$ to $0.0097$~$a_0^{-1}$.
Such a large variation for ethene leads to the observed poor linear correlation.
This discrepancy in the computed excited-state electronic structure causes substantial deviations in the numerator (the non-adiabatic coupling vector) of the derivative coupling.
The combination of this numerator error with significant deviations in the denominator (the excitation energy gap) results in both a poor linear correlation and a slope that deviates from the ideal $y=x$ line (Figure \ref{fig:fssh_ethene_ee}).


The present results suggest that the TDA-ris scheme might not always be a robust approximation for general FSSH applications.
Therefore, we strongly recommend a preliminary validation similar to the tests presented here before applying the TDA-ris method to other systems.
However, for systems where the relevant states and couplings exhibit strong agreement with the reference TDA results (e.g., acetone and hexatriene examined in this study), the TDA-ris method can be employed with greater confidence.

\section{Conclusion}

In this work, we present an implementation of excited-state gradients and derivative couplings for the TDDFT-ris framework based on the standard TDDFT formalism and program infrastructure.
We examine the TDDFT-ris method with respect to standard TDDFT in terms of its computational performance and accuracy for applications involving excited-state potential energy surfaces and excited-state dynamics.

Our benchmarks indicate that for calculations pertinent to excited-state dynamics, the TDDFT-ris approximation provides a moderate enhancement in performance, achieving an overall acceleration of approximately two- to three-fold.
In terms of accuracy, TDDFT-ris generally yields reasonable results for properties such as optimized geometries and excitation energies when compared to standard TDDFT.
However, the inherent errors in the TDDFT-ris excitation energies and electronic structures, relative to TDDFT, can be amplified during the calculation of derivative couplings.
This phenomenon is particularly pronounced in the evaluation of transition probabilities, where significant deviations are observed for certain systems and electronic states. 
Furthermore, the energetic ordering of excited states in TDDFT-ris calculations may differ from that in standard TDDFT, which can complicate the interpretation of results.

Consequently, TDDFT-ris, in its current formulation, is not recommended as a general-purpose, ``black-box'' tool for excited-state dynamics, as its reliable application requires considerable user expertise.
Fortunately, its viability for a specific application can be proactively assessed.
Prior to commencing a full-scale dynamics simulation, one can perform preliminary benchmark calculations on the system of interest---such as determining emission energies, locating MECP, or evaluating time-derivative non-adiabatic coupling matrices---to gauge whether the TDDFT-ris approach exhibits sufficient accuracy for the target calculation.

The computational efficiency and certain functionalities of the TDDFT-ris framework could be further improved, which we plan to address in future works.
First, solving the $Z$-vector equations is the primary bottleneck in derivative calculations.
Applying approximations to the $Z$-vector equations, analogous to those used in TDDFT-ris, could potentially accelerate this step by an order of magnitude without loss of accuracy.
Additionally, TDDFT-ris inherits the well-known limitations of standard TDDFT for computing derivative couplings between the ground and excited states\cite{levine2006conical}.
Approaches such as spin-flip TDDFT\cite{levine2006conical,zhang2021nonadiabatic,herbert2022spin, pu2023noncollinear,li2023noncollinear} could be extended to the TDDFT-ris framework to address these deficiencies.

\section{Acknowledgment}
The author, Zhichen Pu, would like to express his gratitude to Zikuan Wang and Yunlong Xiao for valuable discussions.

\clearpage

\bibliographystyle{plainnat}
\bibliography{references} 

\clearpage

\beginappendix

\section{Appendix: Detailed Formulations for Excited-State Gradients and Derivative Couplings} \label{sec:appendix_formulas}
This section presents the detailed formulations for the excited-state gradients and derivative couplings.

\subsection{Definitions for the Generating Function} \label{subsec:appendix_generating}

The matrices and notations that appear in Eqs.~\eqref{equ:gI} to \eqref{equ:gIJ} are defined in Ref.~\citenum{li2014first} and are reproduced here for convenience.
\begin{align}
    \boldsymbol{T}_{IJ} = & \frac{1}{2} \bigl( \boldsymbol{\gamma}^{IJ}(\mathrm{II}) + \bigl[ \boldsymbol{\gamma}^{IJ}(\mathrm{II}) \bigr]^T \bigr), \label{equ:T} \\
    \quad \boldsymbol{\gamma}^{IJ}(\mathrm{II}) = & \boldsymbol{C} \gamma^{IJ}(\mathrm{II}) \boldsymbol{C}^T, \\
    \gamma_{pq}^{IJ} = & \gamma_{pq}^{IJ}(\mathrm{II}) = \begin{bmatrix}
        -(X_J^T X_I + Y_I^T Y_J)_{ij} & 0 \\
        0 & (X_I X_J^T + Y_J Y_I^T)_{ab}
        \end{bmatrix},  \\
    \boldsymbol{\Gamma}_{IJ} = &  \{ \boldsymbol{R}_I^S, \boldsymbol{R}_J^S \} + \{ \boldsymbol{L}_I^A, \boldsymbol{L}_J^A \}, \\
    \boldsymbol{R}_I^S = & \frac{1}{2} \bigl( \boldsymbol{R}_I + \boldsymbol{R}_I^T \bigr), \quad
    \boldsymbol{R}_I = \boldsymbol{C}_\mathrm{V} R_I \boldsymbol{C}_\mathrm{O}^T, \quad R_I =  X_I + Y_I, \\
    \boldsymbol{L}_I^A =& \frac{1}{2} \bigl( \boldsymbol{L}_I - \boldsymbol{L}_I^T \bigr), \quad L_I =  X_I - Y_I, \label{equ:L_trasition_density} \\
    [\underline{\boldsymbol{d}}]_{pq} = & \langle p | q(x) \rangle, \\
    \quad \underline{\boldsymbol{\gamma}} = & \boldsymbol{C} \gamma [\boldsymbol{C}(x)]^T, \\
    \gamma_{pq}^{0I} = & \begin{bmatrix} 0 & X_{I, ai} \\ Y_{I, ai} & 0 \end{bmatrix}.
\end{align}
Here, boldface letters denote quantities in the atomic orbital basis, whereas plain-type letters represent quantities in the molecular orbital basis.
The underline indicates that only the ket depends on the nuclear coordinate $x$ in atomic-orbital basis. 
Following Ref.~\citenum{li2014first}, the contraction of an integral with density matrices is denoted as
\begin{align}
   \langle O_n; \{ D_1, D_2, \ldots, D_n \} \rangle 
    =  \sum_{P_1 P_2 \cdots P_n} O_{P_1 P_2 \cdots P_n} (D_1)_{P_1} (D_2)_{P_2} \cdots (D_n)_{P_n},
\end{align}
where $P_i$ represents a pair of molecular orbital indices, such as $pq$. 
For example, the contraction of the second term on the right-hand side of Eq. \eqref{equ:gI} reads
\begin{align}
    \langle \boldsymbol{g};  \boldsymbol{\Gamma}_{II} \rangle = & \langle \boldsymbol{g};  \{ \boldsymbol{R}_I^S, \boldsymbol{R}_J^S \} + \{ \boldsymbol{L}_I^A, \boldsymbol{L}_J^A \} \rangle \\
    = & \langle \boldsymbol{g};  \boldsymbol{R}_I^S, \boldsymbol{R}_J^S \} \rangle
        + \langle \boldsymbol{g};  \{ \boldsymbol{L}_I^A, \boldsymbol{L}_J^A \} \rangle \\
    = & \sum_{pqrs} g_{pq,rs} [\boldsymbol{R}_I^S]_{pq} [\boldsymbol{R}_J^S]_{rs} \nonumber 
    + \sum_{pqrs} g_{pq,rs} [\boldsymbol{L}_I^A]_{pq} [\boldsymbol{L}_J^A]_{rs}.
\end{align}
For detailed definitions of these terms, the reader is referred to Ref.~\citenum{li2014first}.

\subsection{Terms for Calculating Derivatives in Standard TDDFT} \label{subsec:appendix_derivatives}

The main text presents the formulas for excited-state gradients and derivative couplings using the notation defined in Ref. \citenum{li2014first} without detailed clarification, which is provided here for completeness.
In Eqs. \eqref{equ:gI_xi} to \eqref{equ:gIJ_xi}, $ \boldsymbol{H} $ is the matrix representing the kinetic energy and the electron-nucleus attraction, and the density-like matrices are defined as
\begin{align}
  \boldsymbol{P}_{I} &= \boldsymbol{T}_{II} + \boldsymbol{Z}^S, \\
  \boldsymbol{Z}^S &=  \frac{1}{2} (\boldsymbol{Z} + \boldsymbol{Z}^T), \quad \boldsymbol{Z} = \boldsymbol{C}_V Z \boldsymbol{C}_O^T, \\
  \boldsymbol{P}_{IJ} &= \boldsymbol{T}_{IJ} + \boldsymbol{\tilde{Z}}^S, \\
  \tilde{M} &= (E_J - E_I) M, \quad M = Z, W, \gamma^{IJ}, L_{IJ}, \\
  \boldsymbol{\tilde{\gamma}}^{IJ} & = \boldsymbol{\tilde{\gamma}}^{IJ}(\boldsymbol{\text{II}}).
\end{align}
The XC potential matrix, $ \boldsymbol{v}_{\text{xc}} $, used in Eqs. \eqref{equ:gI_xi} to \eqref{equ:gIJ_xi}, is defined by its matrix elements as
\begin{align}
  (\boldsymbol{v}_{\text{xc}})_{pq} = & \frac{\partial E_{\text{xc}}}{\partial D_{qp}}.
\end{align}

The matrices $ \boldsymbol{Z} $ and $\boldsymbol{W}$ in Eqs. \eqref{equ:gI_xi} to \eqref{equ:gIJ_xi} need to be determined. 
The matrix $ \boldsymbol{Z} $ is determined from the $Z$-vector equation \eqref{equ:Z_vector},
\begin{align}
    (\boldsymbol{A} + \boldsymbol{B}) \boldsymbol{Z} = g^{(\mathrm{VO})} - g^{(\mathrm{OV})},
\end{align}
with the definition (Eq. (41) in Ref. \citenum{li2014first})
\begin{align}
    L^{(pp')} = \sum_{q} \frac{\partial L}{\partial C_{q p}(x)} C_{q p'}(x).
\end{align}
The matrix element $g^{(pp')}$ adopts different forms for the gradient, the derivative couplings between the ground and excited states, and the derivative couplings between different excited states.
The matrix $\boldsymbol{W}$ is calculated as
\begin{align}
    W_{ij} &= \frac{1}{2} g^{(ij)} + G_{ij}[\boldsymbol{Z}^S], \label{equ:w0}\\
    W_{ab} &= \frac{1}{2} g^{(ab)}, \\
    W_{ia} &= \frac{1}{2} g^{(ia)} + G_{ia}[\boldsymbol{Z}^S] + \frac{1}{2} \sum_{b} F_{ab} Z_{bi}, \\
    W_{ai} &= \frac{1}{2} g^{(ai)} + \frac{1}{2} \sum_{j} Z_{aj} F_{ji}, \label{equ:wf}
\end{align}
with the notation defined as
\begin{align}
    G[\boldsymbol{Z}^S] = \boldsymbol{C}^T \boldsymbol{G}[\boldsymbol{Z}^S] \boldsymbol{C}, \quad \boldsymbol{G}[\boldsymbol{Z}^S] = \boldsymbol{g}[\boldsymbol{Z}^S] + \boldsymbol{f}_{\text{xc}}[\boldsymbol{Z}^S].
\end{align}

For the excited-state gradient, the matrix element $g^{(pp')}$ is calculated as
\begin{align}
  g_I^{(pp')} 
= & 2\bigl[ D  (G[\boldsymbol{T}_{II}] + k_{\text{xc}}[\boldsymbol{R}_{I}^{S}, \boldsymbol{R}_{I}^{S}])\bigr]_{pp'} 
 + 2\bigl[ T_{II}F \bigr]_{pp'} \nonumber \\
& + 4\bigl[ R_I^S G[\boldsymbol{R}_I^S] \bigr]_{pp'} - 4\bigl[ L_I^A g[\boldsymbol{L}_I^A] \bigr]_{pp'}, \label{equ:g_I_pp}
\end{align}
where the definitions for the density matrices, from Eq. (82) in Ref. \citenum{li2014first}, are
\begin{align}
    &D = \begin{bmatrix} I & 0 \\ 0 & 0 \end{bmatrix}, \\
    &T_{II} = \begin{bmatrix} -\frac{1}{2}(R_I^T R_I + L_I^T L_I) & 0 \\ 0 & \frac{1}{2}(R_I R_I^T + L_I L_I^T) \end{bmatrix}, \\
    &R_I^S = \begin{bmatrix} 0 & \frac{1}{2} R_I^T \\ \frac{1}{2} R_I & 0 \end{bmatrix}, \quad L_I^A = \begin{bmatrix} 0 & -\frac{1}{2} L_I^T \\ \frac{1}{2} L_I & 0 \end{bmatrix}, \\
    & \langle g ; \{ D_1, D_2 \} \rangle = \langle g[D_1] ; \{D_2 \} \rangle = \langle g[D_2] ; \{D_1 \} \rangle.
\end{align}
In Eq. \eqref{equ:g_I_pp}, $k_{\text{xc}}$ is defined as the third-order derivatives of the XC functional with respect to the density matrix $\boldsymbol{D}$.

For the derivative coupling between the ground state and excited state $I$, the matrix element $g^{(pp')}$ is defined as
\begin{align}
    g_{0I}^{(pp')} = \gamma_{p'p}^{0I}. \label{equ:g_0I_pp}
\end{align}
For the derivative couplings between excited states $I$ and $J$, the matrix element $g^{(pp')}$ is defined as
\begin{align}
  g_{IJ}^{(pp')} = & (E_J - E_I)^{-1}Q_{IJ}^{(pp')} + \gamma_{p'p}^{IJ} \label{equ:g_IJ_pp}, \\
  Q_{IJ}^{(pp')} = & 2\bigl[ D  (G[\boldsymbol{T}_{IJ}] + k_{\text{xc}}[\boldsymbol{R}_{I}^{S}, \boldsymbol{R}_{J}^{S}])\bigr]_{pp'} + 2\bigl[ T_{IJ}F \bigr]_{pp'} \nonumber \\
  & + 2\bigl[ R_I^S G[\boldsymbol{R}_J^S] \bigr]_{pp'} - 2\bigl[ L_I^A g[\boldsymbol{L}_J^A] \bigr]_{pp'} \nonumber \\ 
  & + 2\bigl[ R_J^S G[\boldsymbol{R}_I^S] \bigr]_{pp'} - 2\bigl[ L_J^A g[\boldsymbol{L}_I^A] \bigr]_{pp'}.
\end{align}

\subsection{Terms for calculating derivatives in TDDFT-ris} \label{subsec:appendix_derivatives_ris}

Within the TDDFT-ris framework, the matrices $\boldsymbol{Z}$ and $\boldsymbol{W}$ are determined according to Eqs. \eqref{equ:Z_vector} and \eqref{equ:w0}-\eqref{equ:wf}, respectively.
However, the matrix element $g^{(pp')}$ on the right-hand side of the $Z$-vector equation, Eq. \eqref{equ:Z_vector}, is defined differently.
For the excited-state gradient, this term is defined as:
\begin{align}
  g_I^{(pp')} 
= & 2\bigl[ D  G[\boldsymbol{T}_{II}]\bigr]_{pp'} 
+ 2\bigl[ T_{II}F \bigr]_{pp'} \nonumber \\
& + 4\bigl[ R_I^S g^{\text{ris}}[\boldsymbol{R}_I^S] \bigr]_{pp'} - 4\bigl[ L_I^A g^{\text{ris}}[\boldsymbol{L}_I^A] \bigr]_{pp'}. \label{equ:g_I_pp_ris}
\end{align}
For the derivative coupling between the ground state and excited state $I$, the matrix element $g^{(pp')}$ is defined as:
\begin{align}
    g_{0I}^{(pp')} = \gamma_{p'p}^{0I}. \label{equ:g_0I_pp_ris}
\end{align}
This expression is identical to the term $g^{(pp')}$ in the standard TDDFT formulation, as shown in Eq. \eqref{equ:g_0I_pp}.
For derivative couplings between excited states $I$ and $J$, the matrix element $g^{(pp')}$ is defined as:
\begin{align}
    g_{IJ}^{(pp')} = & (E_J-E_I)^{-1}Q_{IJ}^{(pp')} + \gamma_{p'p}^{IJ}, \label{equ:g_IJ_pp_ris} \\
    Q_{IJ}^{(pp')} = & 2\bigl[ D  G[\boldsymbol{T}_{IJ}]\bigr]_{pp'} + 2\bigl[ T_{IJ}F \bigr]_{pp'} \nonumber \\
    & + 2\bigl[ R_I^S g^{\text{ris}}[\boldsymbol{R}_J^S] \bigr]_{pp'} - 2\bigl[ L_I^A g^{\text{ris}}[\boldsymbol{L}_J^A] \bigr]_{pp'} \nonumber \\ 
    & + 2\bigl[ R_J^S g^{\text{ris}}[\boldsymbol{R}_I^S] \bigr]_{pp'} - 2\bigl[ L_J^A g^{\text{ris}}[\boldsymbol{L}_I^A] \bigr]_{pp'}.
\end{align}
Notably, the evaluation of $g^{(pp')}$ for the excited-state gradient and the derivative couplings between excited states does not require third-order derivatives of the XC functional, yet it still necessitates the computation of the four-center integrals in $G$.

\section{Appendix: Validation with Finite Differences} \label{app:benchmark_test}
To validate the implementation, the analytical excited-state gradients and derivative couplings are benchmarked against numerical finite difference calculations, for which a step size of $0.001$~\AA\ is utilized.

Table \ref{tab:benchmark_grad} presents the benchmark data for the excited-state gradients.
These results are obtained using both the exact algorithm and the density fitting approximation for the evaluation of two-electron integrals.
The molecular geometries for this test are adopted from Ref.~\citenum{schreiber2008benchmarks}.
For all molecules tested, the analytical gradients show excellent agreement with their finite-difference counterparts.
The MAD and RMS deviations are consistently on the order of $10^{-6}$ to $10^{-5}$ Hartree/Bohr, thus confirming the correctness of the analytical excited-state gradient implementation.

\begin{table*}[htbp] 
  \centering
  \caption{Comparison of analytical and numerical finite-difference gradients for the first excited state (unit in Hartree/Bohr).}
  \begin{tabular}{
    l
    S[table-format=1.2e-2]
    S[table-format=1.2e-2]
  }
    \toprule
    Molecule & {MAD} & {RMS} \\
    \midrule
    acetamide         & 4.53e-06 & 8.35e-06 \\
    acetone           & 4.62e-06 & 7.06e-06 \\
    benzene           & 6.73e-06 & 1.30e-05 \\
    butadiene         & 6.17e-06 & 1.12e-05 \\
    cyclopentadiene   & 7.33e-06 & 1.61e-05 \\
    cyclopropene      & 2.06e-06 & 3.21e-06 \\
    ethene            & 1.26e-06 & 2.52e-06 \\
    formaldehyde      & 1.08e-06 & 2.62e-06 \\
    formamide         & 3.77e-06 & 7.43e-06 \\
    furan             & 1.30e-05 & 2.85e-05 \\
    hexatriene        & 5.66e-06 & 8.92e-06 \\
    imidazole         & 9.14e-06 & 1.84e-05 \\
    pyrazine          & 3.35e-06 & 5.85e-06 \\
    pyridazine        & 1.14e-05 & 1.97e-05 \\
    pyridine          & 3.93e-06 & 8.31e-06 \\
    pyridine          & 9.43e-06 & 1.89e-05 \\
    pyrrole           & 8.91e-06 & 2.02e-05 \\
    tetrazine         & 1.18e-05 & 1.86e-05 \\
    triazine          & 1.55e-05 & 3.31e-05 \\
    \bottomrule
  \end{tabular}
  \label{tab:benchmark_grad} 
\end{table*}

The accuracy of the analytical derivative coupling implementation is benchmarked against the finite difference method, with the results presented in Table \ref{tab:benchmark_nacv}. 
Table \ref{tab:benchmark_nacv} lists the MAD and RMS deviations for the derivative couplings between the ground state ($S_0$) and the first excited state ($S_1$), denoted as $\boldsymbol{g}_{01}^{\xi}$, and between the first ($S_1$) and second ($S_2$) excited states, denoted as $\boldsymbol{g}_{12}^{\xi}$. 
The results demonstrate excellent agreement for both $\boldsymbol{g}_{01}^{\xi}$ and $\boldsymbol{g}_{12}^{\xi}$. 
This high level of accuracy is consistent for calculations employing either the exact two-electron integrals or the density fitting approximation. 
Deviations from the numerical reference consistently fall within the range of $10^{-7}$ to $10^{-4}$~$a_0^{-1}$, which strongly validates the present implementation. 
Although the finite difference method serves as a valuable benchmark, its application to derivative couplings requires particular caution. 
For molecules with degenerate or near-degenerate molecular orbitals, obtaining a consistent difference of the molecular orbital coefficients is non-trivial, rendering the numerical approach challenging\cite{li2014first}. 

\begin{table*}[htbp]
    \centering
    \caption{Comparison of analytical and numerical finite-difference derivative coupling (unit in $ a_0^{-1} $).}
    \label{tab:benchmark_nacv} 
    \begin{threeparttable}
      \begin{tabular}{l S[table-format=1.2e-2] S[table-format=1.2e-2] S[table-format=1.2e-2] S[table-format=1.2e-2] S[table-format=1.2e-2] S[table-format=1.2e-2] S[table-format=1.2e-2] S[table-format=1.2e-2]}
        \toprule
        Molecule & {MAD($ \boldsymbol{g}_{01}^{\xi} $)} & {RMS($ \boldsymbol{g}_{01}^{\xi} $)} & {MAD($ \boldsymbol{g}_{12}^{\xi} $)} & {RMS($ \boldsymbol{g}_{12}^{\xi} $)} \\
        \midrule
        acetamide       & 2.22e-06 & 4.35e-06 & 5.50e-06 & 1.40e-05 \\
        acetone         & 4.98e-06 & 7.83e-06 & 1.13e-05 & 1.81e-05 \\
        cyclopropene    & 1.56e-06 & 3.08e-06 & 2.21e-05 & 4.70e-05 \\
        ethene          & 8.57e-08 & 1.81e-07 & 5.85e-07 & 9.07e-07 \\
        formamide       & 3.90e-05 & 1.07e-04 & 1.23e-05 & 2.82e-05 \\
        propanamide     & 1.33e-05 & 5.42e-05 & 1.98e-05 & 5.96e-05 \\
        \bottomrule
      \end{tabular}
    \end{threeparttable}
\end{table*}

\end{document}